\documentclass[a4paper,final,10pt]{iopart}
\usepackage{iopams}
\usepackage{graphicx}
\usepackage{amssymb}
\usepackage{bm}
\usepackage{color}
\newcommand{\w}{\omega}

\begin{document}
\jl{3}

\topical{Universal transport signatures in two-electron molecular quantum dots:
gate-tunable Hund's rule, underscreened Kondo effect and quantum phase transitions}

\author{Serge Florens$^1$, Axel Freyn$^1$, Nicolas Roch$^1$, Wolfgang Wernsdorfer$^1$, 
Franck Balestro$^1$, Pablo Roura-Bas$^2$, A A Aligia$^3$}
\address{$^1$ Institut N\'eel, CNRS et Universit\'e Joseph Fourier, BP 166,
	38042 Grenoble Cedex 9, France}
\address{$^2$ Centro At\'omico Constituyentes, Comisi\'on Nacional de Energ\'ia At\'omica, 
	1650 San Mart\'in, Buenos Aires, Argentina}
\address{$^3$ Centro At\'omico Bariloche and Instituto Balseiro,
	Comisi\'on Nacional de Energ\'ia At\'omica, 8400 Bariloche, Argentina}

\begin{abstract}
We review here some universal aspects of the physics of two-electron
molecular transistors in the absence of strong spin-orbit effects. Several recent 
quantum dots experiments have shown that an {\it electrostatic} backgate could
be used to control the energy dispersion of {\it magnetic} levels.
We discuss how the generically asymmetric coupling of the metallic contacts
to two different molecular orbitals can indeed lead to a {\it gate-tunable Hund's
rule} in the presence of singlet and triplet states in the quantum dot.
For gate voltages such that the singlet constitutes the (non-magnetic) ground state,
one generally observes a suppression of low voltage transport, which can yet be restored 
in the form of enhanced cotunneling features at finite bias. 
More interestingly, when the gate voltage is controlled to obtain the triplet configuration, 
spin $S=1$ Kondo anomalies appear at zero-bias, with non-Fermi liquid features related 
to the underscreening of a spin larger than 1/2.
Finally, the small bare singlet-triplet splitting in our device allows to
fine-tune with the gate between these two magnetic configurations, leading to 
an unscreening quantum phase transition. This transition occurs between the 
non-magnetic singlet phase, where a two-stage Kondo effect occurs, and
the triplet phase, where the partially compensated (underscreened) 
moment is akin to a magnetically ``ordered'' state.
These observations are put theoretically into a consistent global picture by
using new Numerical Renormalization Group simulations, taylored to capture
sharp finie-voltage cotunneling features within the Coulomb diamonds, together with 
complementary out-of-equilibrium diagrammatic calculations on the two-orbital Anderson model. 
This work should shed further light on the complicated puzzle still raised by multi-orbital 
extensions of the classic Kondo problem.
\end{abstract}
\submitto{\JPCM}

\maketitle
\mathindent=1cm

\twocolumn

\tableofcontents
\makeatletter
\@mkboth{}{Topical Review}
\makeatother

\section{Introduction}

Quantum dots, artificial nanostructures with quantized electronic charge
that can be probed by electrical transport, offer promising ways of 
manipulating the spin of electrons in atomic size devices~\cite{Datta,DiVentra}. By using
a combination of lithographic methods, electrical gates and applied magnetic
field, a high degree of control of single-electron magnetism can be thus
achieved, allowing {\it e.g.} the realization of spin-qubits \cite{Hanson}.
By opening the tunneling barrier between quantum dots and the metallic electrodes
used as contacts, quite fascinating physics emerges at strong tunnel coupling.
In the case of an odd number of trapped electrons in a dot with a single
orbital well separated in energy from other excitations (due to confinement and
Coulomb blockade), an artificial and tunable version of the spin $S=1/2$ Kondo effect 
can be realized~\cite{Hewson,PustilnikGlazman,GrobisReview}. 
In that situation, magnetic screening of the single spin by the Fermi sea occurs
below the so-called Kondo temperature $T_K$, giving rise to a conductance increase up to
the maximum unitary value $2e^2/h$, with $e$ the electron charge and $h$
Planck's constant~\cite{GlazmanRaikh,NgLee}.
The first observation in semiconducting quantum dots of the Kondo effect~\cite{GoldhaberGordon,Cronenwett} 
has triggered intense experimental research to make similar observations in
other physical systems such as carbon nanotubes~\cite{Jespersen2006} and
molecular devices~\cite{Liang,ParksKondo}, see Ref.~\cite{NatelsonReview} for a review.
In parallel, many theoretical works have followed, and at present reliable methods, 
such as the Numerical Renormalization Group (NRG) \cite{Wilson,BullaRMP}, have been
developped. These calculations can allow quantitative comparison to experiments, especially 
because the spin $S=1/2$ Kondo effect shows simple universal scaling laws, that
can be checked in quantum transport measurements.
At present, one important goal is to understand in similar detail the electronic transport 
through multi-orbital dots, a situation clearly raised both by semiconducting
and molecular devices.
The next step towards complexity, two-electron quantum dots, seems deceptively
small, yet brings a richness that is still not fully clarified.
Theoretically, a multitude of new effects have been predicted over the years in the 
realm of Kondo physics, and many have been observed thanks to the great
tunability of quantum dot systems. We briefly give an overall review of this physics 
in the rest of this introduction, with additional details provided in Section~\ref{instabilities}.

Let us first focus on the case where the two orbitals experience strong ferromagnetic 
Hund's rule, so that the triplet configuration constitutes the main low lying states.
This can result in the spin $S=1$ underscreened Kondo effect if a {\it single} screening
channel is active, as initially proposed by Nozi\`eres and Blandin~\cite{NozieresBlandin},
and only recently experimentally observed~\cite{RochUS,ParksUS}. This rather
exotic Kondo effect shows a singular (logarithmic) approach to strong coupling,
in contrast to the regular Fermi liquid behavior of fully screened Kondo impurities 
observed in odd-charge spin $S=1/2$ quantum dots \cite{GrobisReview}.
Because Hund's rule competes with the orbital level spacing, an intra-orbital
singlet state is usually close (and often lower) in energy to the triplet states, 
and the competition between singlet and triplet can give rise to very rich physics.

One interesting example is the so-called singlet-triplet ``transition'', which realizes
a fully screened Kondo effect by bringing into degeneracy singlet and triplet 
magnetic levels  (it is thus rather a crossover than a real phase transition),
for which two different scenario can be obtained.
The simplest situation occurs in carbon nanotubes (or molecular quantum dots),
where a strong Zeeman effect is used to cross the singlet state with the {\it lowest} 
triplet state in the presence of a {\it single screening channel}~\cite{Nygard,PustilnikAvishai},
showing the emergence of a strong Kondo enhanced conductance at the crossing point.
Alternatively, similar physics can be realized in semiconducting vertical quantum dots,
where the magnetic field can be used to tune by orbital effects the splitting between
the singlet and the {\it three-fold degenerate} 
triplet~\cite{Sasaki,STKondo_Eto,STKondo_Pustilnik1,STKondo_Pustilnik2}
(the Zeeman effect can be neglected due to the small g-factor). Here 
the vertical structure preserves the orbital quantum numbers during the tunneling 
process, so that two screening channel are active, and the high spin states are 
again fully Kondo screened at the crossing point.

One can then naturally wonder about the fate of a singlet to {\it degenerate} triplet
crossing in the presence of a {\it single} screening channel 
\cite{Allub,VojtaBulla,HofstetterSchoeller,PustilnikSingletTriplet,HofstetterZarand,ZitkoSTMultidot,ZitkoSTTransport,ZarandST,Roch}, 
which can in principle be expected for lateral quantum dots \cite{VanDerWiel,Kogan} and
molecular devices \cite{Roch}.
Interestingly, this situation gives rise to both underscreening physics~\cite{NozieresBlandin,RochUS} 
and an unscreening singlet-triplet {\it quantum phase transition} (QPT).
In that case, a zero-temperature phase transition can take place between a non-magnetic 
singlet state and a partially compensated spin $S=1$ (giving an remanent spin
$S=1/2$ in the ground state). The transition is really sharp at zero temperature
as the entropy changes from zero (in the singlet phase) to $\log(2)$ in
the underscreened Kondo phase. This is in contrast to the so-called singlet-triplet
``transition'' induced by a magnetic field, where the entropy vanishes in both
``phases'' (as discussed in the previous paragraph).
Several different theoretical proposals have also been made for the unscreening
QPT~\cite{HofstetterSchoeller,ZitkoSTTransport,PustilnikBorda,Logan}, which was shown to be even 
robust to valence fluctuations.
Our detailed experimental observation \cite{Roch} of this singlet-triplet quantum phase transition in
a two-electron molecular quantum dot will be one of the main topics of this
review, and the various scenarios presented above will be reviewed with more
details in the concluding Section~\ref{instabilities}.

Out-of-equilibrium Kondo phenomena have also been investigated in two-electron
quantum dots, such as Kondo-enhanced cotunneling lines at finite bias
\cite{Paaske,Roch,RochJLTP}, and these will be mentioned later on.
We stress also that spin-orbit interaction has been given some recent attention both experimentally
\cite{DeFranceschi,Jespersen,ParksUS} and theoretically
\cite{ZitkoSO,Galpin,Tagliacozzo,ParksUS,Cornaglia,Andergassen} 
for even-charge quantum dots, as well as for odd-electron quantum dots
\cite{PaaskeSO}, and these effects could possibly play an important role for the interpretation 
of some experiments made in semi-conducting quantum dots \cite{VanDerWiel,Kogan}.
We also remark that carbon nanotube quantum dots can show a wealth of interesting Kondo 
states due to the enhanced SU(4) symmetry, both in single and doubly occupied situations, 
but these system go beyond the present discussion (see e.g. Ref.~\cite{Jarillo,SU4Choi} 
and references therein).

A different physical situation occurs when the two orbitals interact with antiferromagnetic 
exchange. In the case of a single screening channel, a two-stage Kondo effect
occurs for large Kondo coupling compared to the interspin exchange.
Indeed, after complete Kondo screening of the first spin, the resulting Fermi
liquid state is able to absorb the second spin in a second stage of screening
\cite{Chung2}.
The situation of two screening channels leads potentially to even more exotic
physics. This provides a realization of the so-called two-impurity Kondo problem 
\cite{VarmaJones}, which realizes a competition 
between Kondo screening and (RKKY) exchange and presents a non-trivial quantum
phase transition. The experimental observation of this transition turns out to be
difficult \cite{Chung,Vavilov} despite previous attempts \cite{Craig}, due to
relevant physical processes at the non Fermi liquid fixed point (essentially the interdot tunneling). 

Finally, it is also worth noting the complicated role of charge when capacitive coupling 
between two dots (or two orbitals) is relevant \cite{KoenigGefen}, which can give rise
to complex charge skipping in the filling scheme of the dots, a situation not
fully elucidated in the presence of spinful electrons. 

On a mathematical level, all the important complexity of these various phenomena can be expressed 
in complete generality by a two-orbital Anderson impurity model, which contains a great 
deal of physical parameters compared to the single orbital Anderson or Kondo
Hamiltonians. One crucial question is thus to understand what classes of universal 
behaviors can be expected in such a complicated situation.
Going one step further, we also mention that three-orbital models show even more complex and 
interesting physics, such as non Fermi liquid fixed points, a topic that goes far beyond the present
review (see Ref.~\cite{LeoFabrizio,Lazarovits,Ingersent,AligiaTrimer,MitchellLogan,ZitkoSTMultidot} 
for a few recent theoretical studies and Ref.~\cite{Ferrero} for a review).

On an experimental point of view, multi-orbital effects are always present,
but their relevance will depend on many factors, most primarily the energy 
spacing between the two relevant orbitals, and the structure of the hybridization
matrix. It is clear that disentangling the mechanisms at play from transport 
measurements only is not an easy task, yet many useful information can be
gathered from finite bias cotunneling spectroscopy, possibly in the presence
of magnetic fields. The present study will illustrate this general strategy,
and show how the magnetic states involved can be identified through an analysis
of Zeeman split cotunneling lines.

The paper is organized as follows. Section \ref{exp} will present recent
observations made with fullerene quantum dots obtained by electromigration.
These will concern finite bias cotunneling spectroscopy of magnetic states, 
gate-dependent tuning of singlet and triplet states, the spin $S=1$ underscreened Kondo 
effect, and the unscreening quantum phase transition. Connection to related results 
in both molecular and semiconducting nanoelectronics will also be made, showing the wide range
of application of those ideas. Section \ref{Theo} will discuss modelization
of the problem, and the combined use of near-equilibrium Numerical
Renormalization Group calculation and diagrammatic methods in the non-equilibrium
situation to reproduce semi-quantitatively most of the features seen in the experiment. 
Finally section~\ref{instabilities} will close the paper by presenting a more detailed 
review of the various interesting Kondo phenomena involving singlet and triplet excitations 
in two-electron quantum dots, that will set our observations in a broader perspective.

\section{Experiments in even charge quantum dots: gate control of spin states
and underscreened Kondo effect} 
\label{exp}

Kondo physics has been widely reported in a variety of quantum dot systems,
based on semiconducting heterostructures~\cite{GoldhaberGordon,Cronenwett},
carbon nanotubes~\cite{Jespersen2006} and molecular
devices~\cite{Liang,ParksKondo,Roch,RochUS,NatelsonReview}, showing the
great universality of this phenomenon. Usually the Kondo effect is best observed
for odd charge quantum dots, because the formation of a net spin $S=1/2$ is
always guaranteed in that case (for even charge dots, the ground state turns out often 
to be a non-magnetic singlet). While Kondo signatures associated to a triplet 
configuration in two-electron quantum dots was previously
observed~\cite{Kogan}, little attention was devoted to the detailed study
of this phenomena. One reason is that the Kondo temperature for spin $S=1$
is generically quite low, making semiconducting systems not suitable to explore
this physics. In contrast, molecular devices display much bigger energy scales,
allowing to investigate in great detail the richness of the Kondo effect in
two-electron quantum dots.
The aim of this paper is to expose the generic and if possible universal
features in such two-electron systems, both at the light of experiments on fullerene 
transistors and also thanks to state-of-the-art many-body simulations. This section will be 
devoted to the presentation and discussion of the experiments, while the theory
will be presented in Sec.~\ref{Theo}.

\begin{figure}
\includegraphics[scale=0.32]{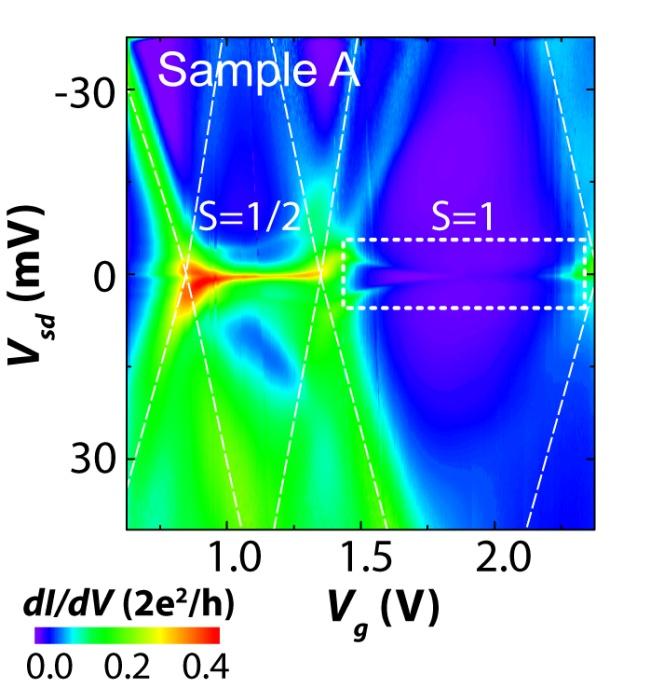}
\includegraphics[scale=0.94]{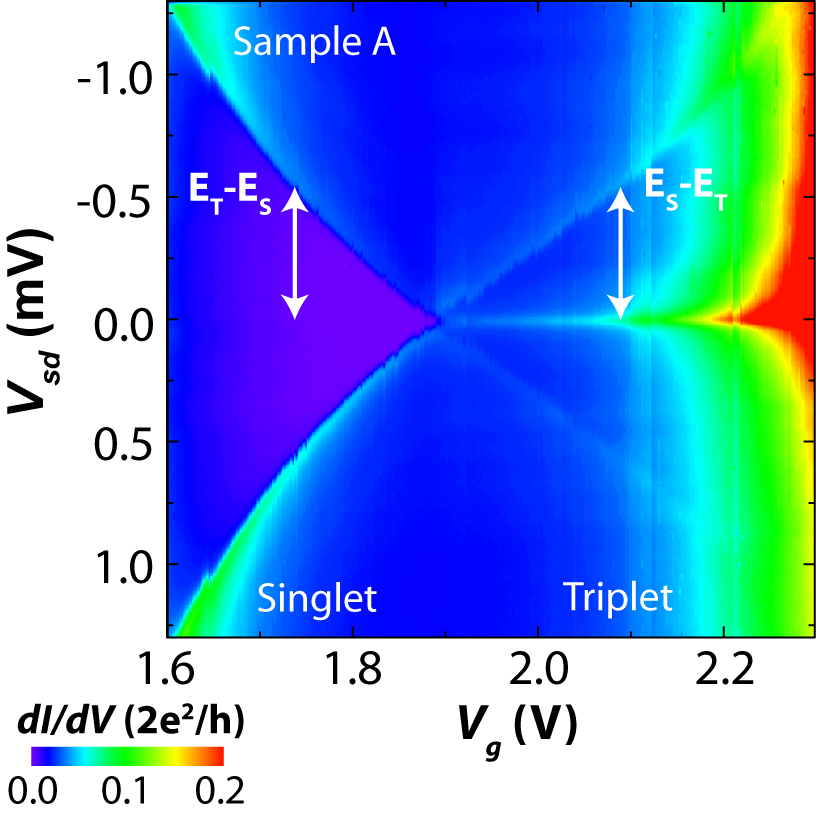}
\vspace{0.1cm}\\
\caption{Observation of a Kondo screening-unscreening transition in an
even charge Coulomb diamond of the molecular transistor \cite{Roch}. The lower
panel is a close up inside the charge $N=2$ Coulomb diamond seen on the upper
panel.
Singlet/triplet states of the molecule are determined from the
absence/presence of a zero bias Kondo anomaly, as well as by the Zeeman 
spectroscopy analysis shown in Fig.~\ref{zeeman}.}
\label{diamond}
\end{figure}

In our previous experimental work \cite{Roch}, a gate-tuned transition from
a low zero-bias conductance state to a high zero-bias conductance 
state was reported in a molecular quantum dot containing
two electrons, see Fig.~\ref{diamond}. The data also show a striking 
collapse of the magnetic excitations on {\it both} sides of the transition,
and can thus naturally be explained by the so-called singlet-triplet unscreening quantum
phase transition
\cite{Allub,VojtaBulla,HofstetterSchoeller,PustilnikSingletTriplet,HofstetterZarand,ZitkoSTTransport}, 
where the low conductance regime (left part of lower panel in Fig.~\ref{diamond}) corresponds to 
a singlet state binding of the molecular orbitals, while the high conductance regime 
(right part of lower panel in Fig.~\ref{diamond}) is associated to the underscreened Kondo 
effect of a spin $S=1$ triplet.
We note that similar features of Kondo anomaly collapse have been reported in 
other two-electron quantum dot systems~\cite{Kogan,Quay,Delattre,ParksUS}, but did not
show a concomitent crossing of the magnetic excitations. Recent theoretical work
\cite{ParksUS,Tagliacozzo} proposed spin-orbit effects (or more generally magnetic
anisotropies) as a possible alternative interpretation for some of these other experiments. 
We will show however that spin-orbit coupling can be neglected in our particular experiment, which is a
crucial requirement \cite{ParksUS,Tagliacozzo} for the robustness of our observation
of Kondo underscreening \cite{RochUS}. Finally, recent experimental 
and theoretical studies \cite{Holm,Hauptmann,Osorio,RochUS,Aligia1,Aligia2,Freyn2} have confirmed several
microscopic mechanisms at play in the singlet-triplet quantum phase transition,
such as the gate tunable Hund's rule, giving extra strength to our initial interpretation. 
We will refer the reader to our previous work\cite{Roch,RochJLTP,RochPhys,RochUS} for 
the experimental details, and will present in what follows a physical discussion of 
our observations.

\subsection{Cotunneling spectroscopy: nature of the magnetic ground state 
and excitations}
\label{spectro}

Identifying the molecular magnetic states involved on the quantum dot is 
certainly a key step in the further interpretation of our transport experiments.
Because Kondo physics results at relatively large tunnel coupling to the electrodes, 
sequential tunneling lines, occuring along the edges of the Coulomb diamonds, are typically 
very broad and masking the useful information on the allowed transitions from a charge 
state to the next. For this reason, cotunneling spectroscopy done within the
Coulomb diamonds (hence at fixed charge on the quantum dot) turns out to be a very useful tool.
Because cotunneling spectroscopy reveals transition lines between states with the
same charge, their intensity is much weaker than transition with a valence
change occuring at the edges of the Coulomb diamonds. This strategy has clear
advantages for improving the spectral resolution, but requires at the
same time careful experimental measurements, performed at subKelvin temperatures
and using efficient filtering of the electric noise.

\begin{figure}
\includegraphics[scale=0.65]{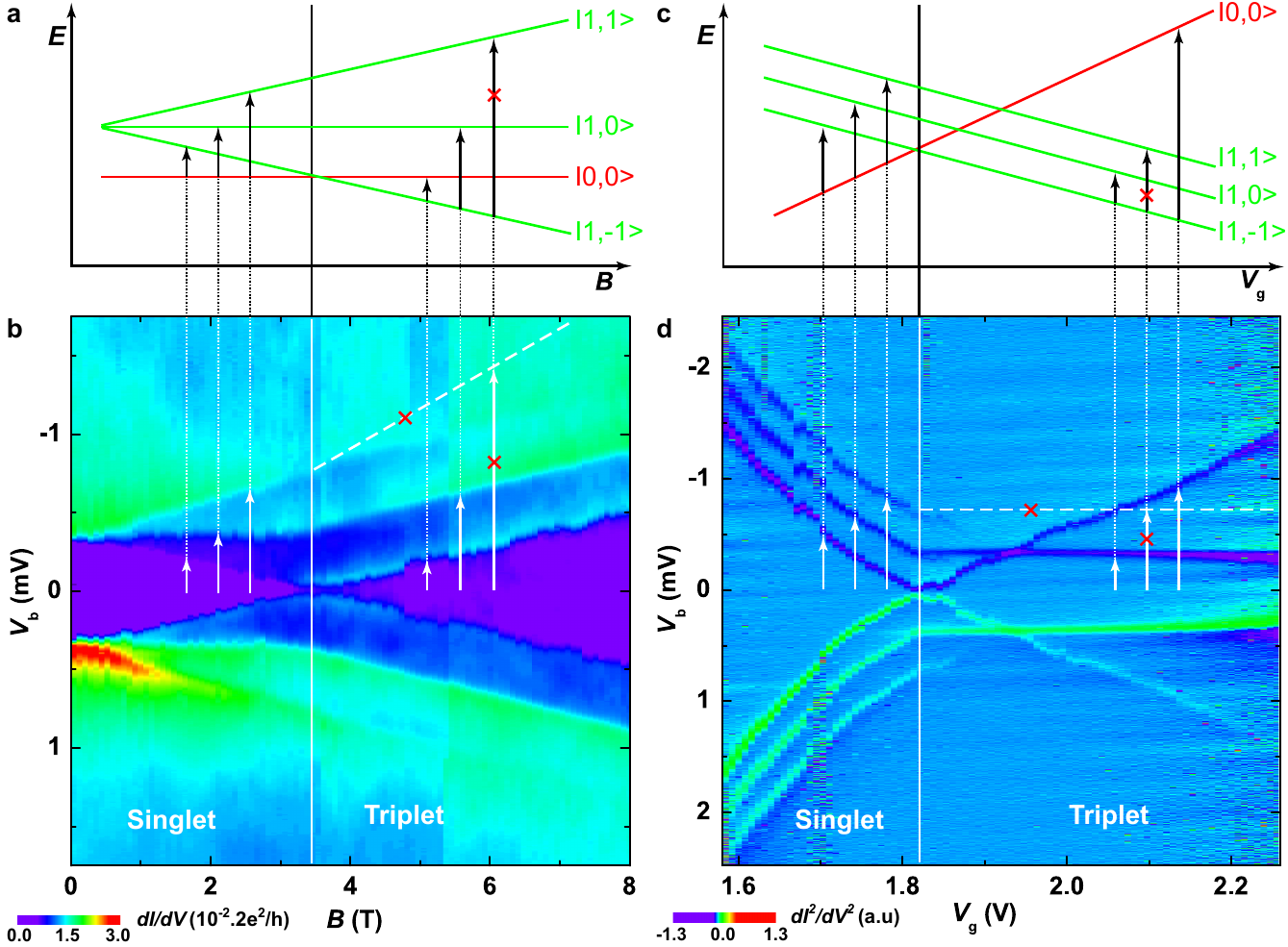}
\caption{Identification of the singlet and triplet excitations by
the Zeeman effect in our device \cite{Roch}. The left panel shows
the magnetic field dependence of the cotunneling lines at a fixed
gate voltage taken in the singlet region. The right panel presents
a gate voltage scan for a given value of the magnetic field (3 Tesla). 
In both cases, spin transitions follow the selection rules associated
to a singlet-triplet scenario.}
\label{zeeman}
\end{figure}

Let us thus focus on the cotunneling lines seen in the even charge Coulomb
diamond of our experiment (lower panel of Fig. \ref{diamond}), and study finite 
bias spectroscopy in a magnetic field. If a spin triplet is really the
ground state for large positive gate voltage, as is hinted by the presence
of a zero-bias anomaly, it should be split in an obvious manner by the Zeeman 
effect, and magnetic transitions should appear accordingly. 
The transport data shown on the right panel of Fig.~\ref{zeeman} clearly
confirms this interpretation, as the selection rules from the ground triplet 
states to the excited singlet state are precisely the ones expected.
Similarly, magnetic-field induced transition from the singlet to the lowest 
triplet is clearly evidenced on the Zeeman lines obtained as a function of the
magnetic field at fixed gate voltage (left panel of Fig.~\ref{zeeman}). 
This understanding of which states are involved in the experiment provides already useful 
information for the interpretation and modelization presented in the following.
Also, the absence of anticrossings between the magnetic excitations shows
that singlet and triplet remain pure spin states. This points to the absence
of spin-orbit effects (see Ref. \cite{DeFranceschi,Jespersen} where such effects
are evidenced in a self-assembled semiconducting quantum dot and carbon
nanotubes respectively), which is for instance a crucial point for the robustness 
\cite{ParksUS,Tagliacozzo} of an underscreened Kondo effect (discussed in section~\ref{underscreened}).

\subsection{Driving mechanism for the gate-controlled singlet-triplet transition}
\label{mechanism}

One striking question should arise at the light of the above analysis:
why a magnetic transition can be driven by a change of the electrostatic
potential? Such phenomena, not fully elucidated in our original work \cite{Roch},
has received a lot of experimental and theoretical attention recently
\cite{Holm,Hauptmann,Osorio,Aligia1,Aligia2} and has a simple explanation in terms 
of molecular levels renormalized by virtual charge fluctuations.

A crucial piece of this puzzle was infact already given in an observation made in
our previous work \cite{Roch}, namely that under a magnetic field the crossing
of the singlet $\big|0\big>$ and the lowest $\big|1,-1\big>$ triplet does not lead to an enhanced
conductance (as seen in the left panel of Fig. \ref{zeeman}), in stark constrast
to the (single-channel) singlet-triplet ``transition''
scenario~\cite{Nygard,PustilnikAvishai}. This implies that, 
at the level crossing, the degenerate ``doublet'' \{$\big|0\big>$, $\big|1,-1\big>$\} 
induced by the magnetic field possesses a vanishingly low Kondo temperature. This 
necessarily means that the two molecular orbitals $(1,2)$ must have an {\it asymmetric tunneling 
amplitude} to the leads, namely $t_1\gg t_2$, since the effective Kondo exchange interaction 
associated to this pair of states is proportional to $t_1 t_2/U$ (with $U$ the 
dot charging energy). Indeed, a ``spin-flip'' between the singlet and lowest
triplet configurations involve tunneling out of orbital 1 and tunneling into
orbital 2 (and vice-versa).

\begin{figure*}
\centerline{\includegraphics[scale=0.20]{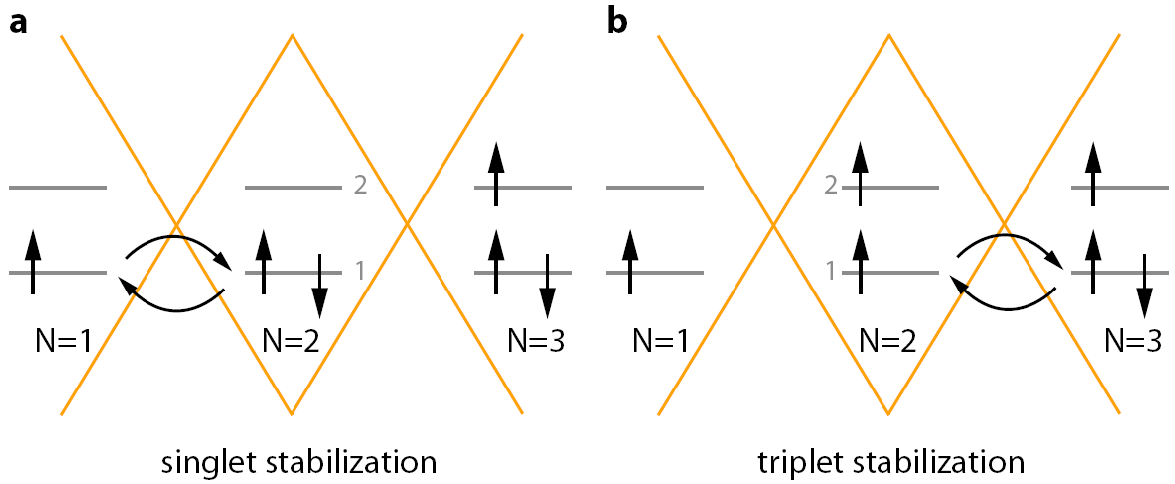}}
\caption{Panel a: stabilization of the singlet state on the left
side of the $N=2$ diamond via virtual fluctuations to the $N=1$ diamond.
Panel b: stabilization of the triplet state on the right
side of the $N=2$ diamond via virtual fluctuations to the $N=3$ diamond. The
mechanism is effective when level $1$ is more strongly hybridized than level $2$
to the leads. See Refs.~\cite{HofstetterSchoeller,ZitkoSTMultidot} for a
detailed theoretical treatment by perturbation theory in the tunneling.}
\label{scheme}
\end{figure*}

Now a qualitative argument, analogous to the one proposed in Refs. 
\cite{Holm,Hauptmann} for odd charge quantum dots, and based on detailed perturbative 
calculations in the tunneling amplitude~\cite{HofstetterSchoeller,ZitkoSTMultidot}, shows 
that the existence of this orbital tunneling asymmetry is the key to the gate-controlled
singlet-triplet splitting. The basic idea relies on a lowering of the singlet
(resp. triplet) energy on the left (resp. right) side of the $N=2$ Coulomb diamond 
by virtual charge fluctuations to the $N=1$ (resp.\ $N=3$) neighboring Coulomb diamond. 
This is due to the fact that the first orbital has indeed a higher probability of tunneling 
than the second one (see Fig. \ref{scheme}), so that an energy gain of order
$(t_1^2-t_2^2)/U$ can be achieved by either the singlet or the triplet
configuration, depending on the proximity to the left or right Coulomb diamond
respectively.
This hypothesis is in agreement with our previous observation~\cite{Roch} of asymmetric
tunneling amplitudes, as discussed above, and this simple mechanism explains 
how a gate-tunable Hund's rule can be achieved in our device.

\begin{figure*}
\centerline{\includegraphics[scale=0.7]{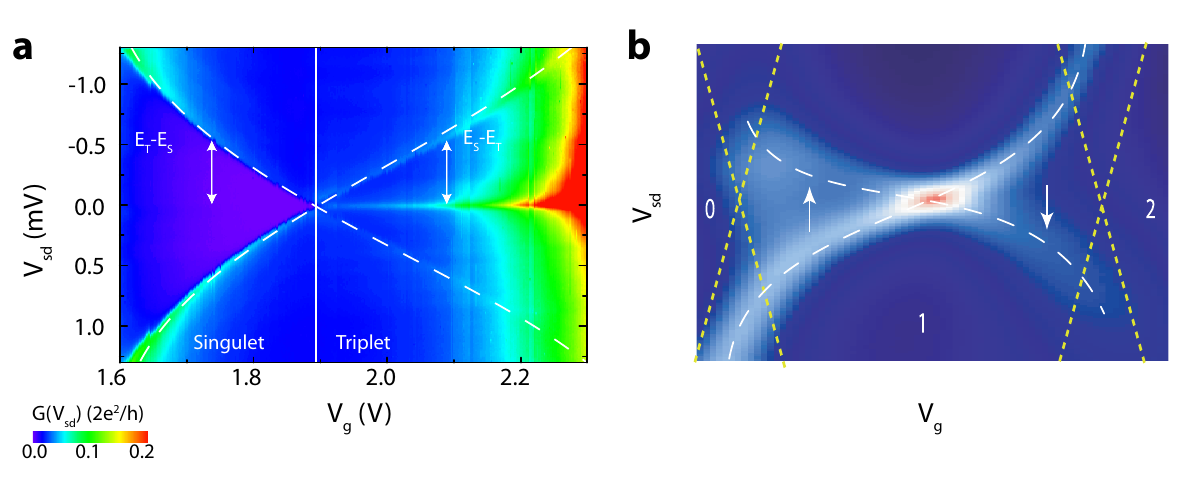}}
\vspace{1.0cm}
\caption{Illustration of the gate-tuning of magnetic states and the bending of
the cotunneling lines at the diamond edges. Panel a: our single
molecule transistor data for an even charge $N=2$ quantum dot \cite{Roch} where a
gate-tunable Hund's rule is achieved. 
Panel b: carbon nanotube device by Hauptmann {\it et al.} \cite{Hauptmann} for
an odd charge $N=1$ quantum dot coupled to ferromagnetic leads, where a gate-tunable
Zeeman splitting is realized.}
\label{gatetuned}
\end{figure*}

On a more quantitative level, one can also understand why the dispersion of
the magnetic states becomes non-linear near the edges of the
Coulomb diamond, as seen in Fig. \ref{gatetuned}. Lowest order perturbation theory 
in the tunneling events associated to the first orbital provides a correction to
the bare singlet-triplet splitting, that diverges logarithmically near the 
diamond edge \cite{Hauptmann}. This sharp enhancement is clearly responsible for the
bending of the magnetic excitation lines near the edge of the Coulomb diamonds,
and should be expected to be a generic effect. Indeed, such features have now 
been widely reported, and for illustration purposes we present data on carbone 
nanotube quantum dots \cite{Hauptmann}, taken in the {\it odd} charge sector 
for a device with {\it ferromagnetic} electrodes, see Fig. \ref{gatetuned}b. 
There, the tunneling asymmetry concerns rather the up and down states of the 
impurity spin $S=1/2$ (so that e.g.  $t_\uparrow\gg t_\downarrow$), and in a 
similar fashion to our findings, leads here to a gate-tuned Zeeman effect for 
these magnetic excitations (the non-linearity near the diamond edges is also 
quite apparent in this data set).

One can thus infer that a gate-induced singlet-triplet transition
can be considered a general effect that will however be observed according to some
quantitative criterium: the bare singlet-triplet splitting of the isolated
molecule (that results from the competition of level spacing, Coulomb repulsion and 
Hund's rule) must be smaller than the singlet to triplet energy dispersion 
that is introduced by the asymmetric coupling to the electrodes. The
singlet-triplet transition can thus only occur at relatively strong
hybridization to the leads.
In alternative cases where the singlet-triplet splitting turns out to be larger
than the hybridization shifts, the quantum dot keeps intact its magnetic ground state 
(singlet or triplet) throughout the whole Coulomb diamond. This is for instance
routinely seen in less strongly coupled quantum dot systems, such as carbon
nanotubes \cite{Paaske}, or our own measurements in other C$_{60}$ 
devices \cite{RochUS}.

\subsection{Triplet side: spin $S=1$ underscreened Kondo effect}
\label{underscreened}

If one focuses on the side of the transition where a Kondo ridge is visible
(large positive gate voltage), a clear spin $S=1$ Kondo anomaly is observed.
One interesting question is the nature of this Kondo state, which will crucially
depend on the number of active screening channels (at the temperature where
the experiment is performed). In the case of two screening channels, the spin
$S=1$ is fully quenched~\cite{PustilnikSingletTriplet}, resulting in a conventional 
Fermi liquid ground state, while for a single screening channel, the spin $S=1$ is
partially compensated. This underscreening process is associated to logarithmic
deviations to Fermi liquid theory \cite{NozieresBlandin}. We note that Kondo
anomalies with even charge states where previously observed, although quite
rarely \cite{Tagliacozzo}, in some other semiconducting quantum dot experiments
\cite{Schmid,Granger}, and were not studied in great detail.

Here, we argue that molecular quantum dots will generically end up in a single 
screening channel situation, in the sense that the Kondo coupling to a second 
screening bath, although always present, will in general be relatively small
(compared to the most strongly coupled channel), so that the associated Kondo temperature 
for a second stage of full screening of the impurity spin will be exponentially small. 
Several arguments are in favor of this situation: i) tunneling from the electrodes 
is typically monomode because the electromigration process is stopped at
the breaking up of the metallic contacts;
ii) coupling of the molecule to the source and drain is also typically quite 
asymmetric (conductance maxima of Kondo anomalies in molecules range from a few 
per-thousands to half the conductance quantum \cite{ParksKondo}); iii) the two molecular 
orbitals involved can also be asymmetrically coupled to the leads, as
is seen at least in our device, see discussion in section~\ref{mechanism}.

\begin{figure}
\includegraphics[scale=0.50]{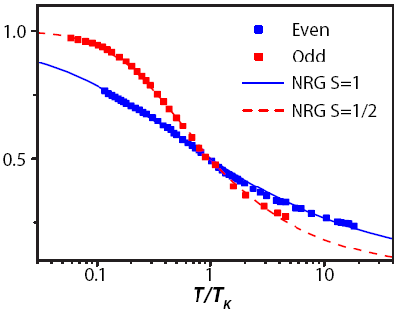}
\caption{Comparison of the (dimensionless) conductance measured in odd and even
Coulomb diamonds (squares) to NRG calculations (lines) \cite{RochUS}, in units
of the Kondo temperature $T_K$.}
\label{US}
\end{figure}

To explain this more presicely, let us introduce the matrix of tunneling 
amplitudes~\cite{PustilnikSingletTriplet}
\begin{equation}
\label{tunnel}
\mathbf{t}=
\left(
 \begin{array}{cc}
t_{L1} & t_{R1} \\
t_{L2} & t_{R2} 
\end{array}
\right),
 \end{equation}
where we describe our molecular transistor by two orbital levels $(1,2)$ coupled 
to two metallic leads $(L,R)$ with single electronic mode.
A screening channel ($\lambda =$even,odd) is associated to each eigenvalue
$t_\lambda$ of this matrix, from which antiferromagnetic Kondo
couplings $J_\lambda=4|t_\lambda|^2/E_\mathrm{add}$ result between the localized
orbitals and the conduction electrons ($E_\mathrm{add}$ is a measure of
the addition energy on the molecule). Typically the even combination of the
wavefunctions results in a larger tunneling term than the odd one, namely 
$t_\mathrm{even}>t_\mathrm{odd}$, because of generic source/drain ($L/R$) asymmetries, 
or also orbital level ($1/2$) asymmetries. In turn, the two exchange interactions 
will show relatively different magnitudes, $J_\mathrm{even} > J_\mathrm{odd}$,
so that a first stage of Kondo underscreening of the $S=1$ triplet will occur at 
temperatures $T_K^{\mathrm{even}}\propto \exp(-1/\rho_0 J_\mathrm{even})$, 
with $\rho_0$ the leads density of states. 
A second stage of screening of the remanent doublet state will develop only at
the much lower temperature $T_K^{\mathrm{odd}}\propto \exp(-1/\rho_0
J_\mathrm{odd})\ll T_K^{\mathrm{even}}$, owing to the exponential 
dependence of the Kondo scale in the exchange interaction. This implies that spin $S=1$ 
molecular quantum dot experiments covering a couple of decades in temperature will only
show underscreening, unless fine-tuning of the tunneling matrix is realized.

This discussion can be put onto firm ground by making quantitative comparisons
of the measured temperature-dependent conductance to Numerical Renormalization Group
calculations, see Fig.~\ref{US} and Ref.~\cite{RochUS} for further
details. The slow saturation of the low temperature conductance for the even
charge spin $S=1$ regime, associated to the logarithmic approach to the 
underscreening fixed point, is readily contrasted to the rapid Fermi liquid-like
behavior at low temperature for the odd charge spin $S=1/2$ Kondo effect. This
{\it quantitative} comparison gives also more convincing evidence for our
observation~\cite{Roch} of a quantum phase transition when singlet and triplet
states meet at the critical gate voltage.

\subsection{Singlet-triplet unscreening quantum phase transition}

The sudden disappearance of the underscreened Kondo resonance in the middle of the
even charge Coulomb diamond (see lower panel in Fig.~\ref{diamond}) is the sign of 
a possible quantum phase transition between two different sorts of ground states.
From the identification of the singlet and triplet magnetic states 
performed in section~\ref{spectro}, and the confirmation that a single 
screening channel is active (see discussion in sections~\ref{mechanism}
and~\ref{underscreened}), 
the obvious scenario to follow is the so-called singlet-triplet unscreening
transition~\cite{Allub,VojtaBulla,HofstetterSchoeller,ZitkoSTMultidot,ZitkoSTTransport}.
In this picture, the transition is driven by the merging of singlet and 
(underscreened) triplet magnetic excitations together at a quantum critical 
point. The transition is continuous, but shows an entropy change at the 
zero-temperature critical point: the non-degenerate singlet state cannot be 
smoothly related to a partially screened spin with $\log(2)$ remanent entropy.
Because the zero-temperature limit where the transition takes really place is 
unaccessible to experiments, we discuss the associated signatures of this 
transition at finite temperature. We already note that the physics here is
more complex than a simple singlet/triplet (or singlet/doublet) level crossing, 
because of the electronic correlations arising from the electrodes, and cannot
be described by a first order transition.  
\begin{figure}
\includegraphics[width=9cm]{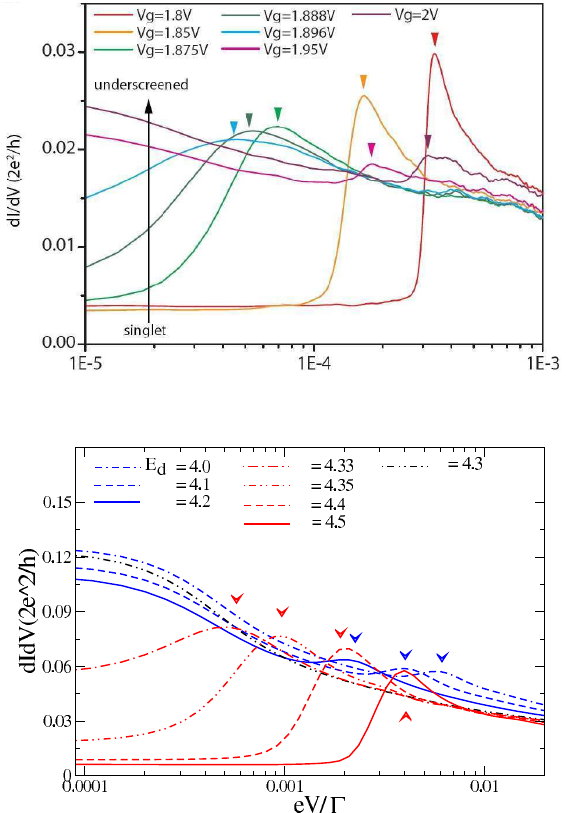}
\caption{Differential conductance as a function of bias voltage for
several gate voltages across the transition. 
In both plots, arrows denote the dispersive singlet/triplet excitations. 
The upper panel shows the experimental measurement \cite{Roch} and the lower panel the 
theory~\cite{Aligia1,Aligia2} performed in section~\ref{diagram}. 
The theoretical calculations are performed for several values of $E_d$ across 
the QPT, with parameters $D=20$, $E_t=0$, $E_s=-0.04$, $\Gamma_1=1.106$, 
$\Gamma_2=0.38$, and $T=10^{-4}$.
In the experimental plot (upper panel), the four lower curves (associated to low conductance 
at small bias) are taken on the singlet side, while the upper curves correspond to the triplet side. 
}
\label{didv}
\end{figure}

\begin{figure}
\includegraphics[scale=0.20]{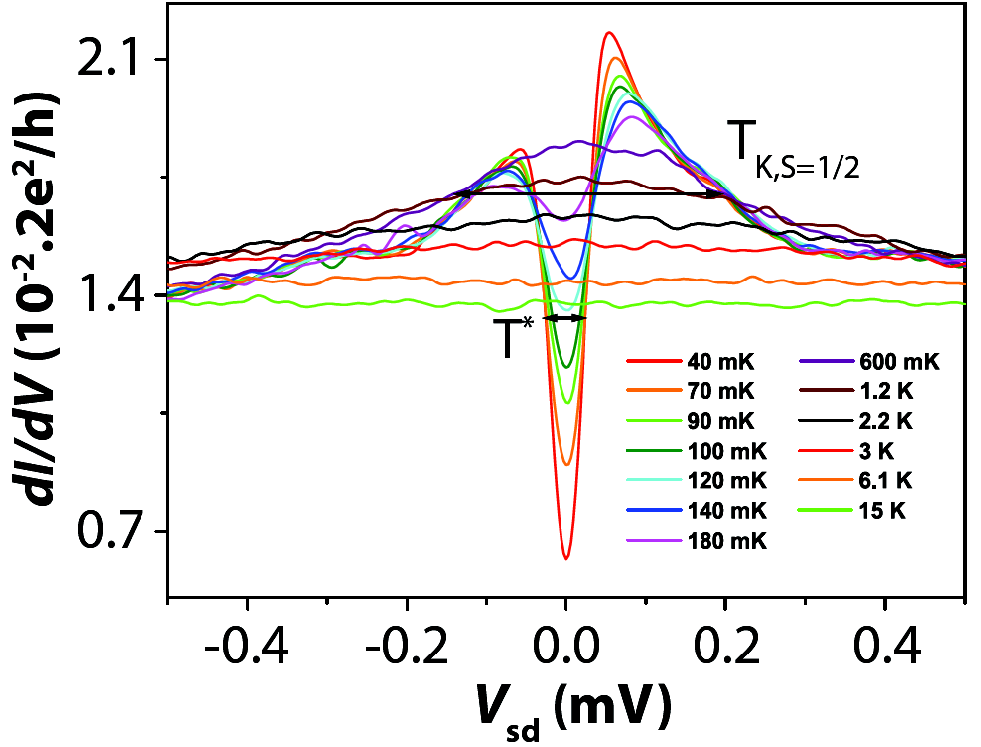}
\caption{Differential conductance on the singlet side of the transition for
several temperatures from 40mK to 15K, showing the two-stage kondo process.}
\label{SingletSide}
\end{figure}

The upper panel of figure~\ref{didv} shows previously unpublished results for the non-linear
conductance measured for several gate voltages across the transition. Far on
the singlet side (for gate voltages $V_g<$1.87V), a well formed gap opens up in
transport, showing the strong binding of the molecular levels into a tight singlet unit. 
Yet an out-of-equilibrium enhancement of the cotunneling lines (to the triplet excitations) 
is witnessed~\cite{RochPhys}, similar to previous observations made in carbon
nanotubes~\cite{Paaske}. 
Upon approaching the transition (for gate voltages 1.88V$<V_g<$1.9V), the finite bias shoulder 
progressively disperses towards zero bias, while a slower logarithmic-like decrease of the conductance 
is now observed instead of a hard singlet gap. These features can be explained within the singlet-triplet 
scenario \cite{Allub,VojtaBulla,HofstetterSchoeller,ZitkoSTTransport}
by a two-stage Kondo process for the formation of the singlet ground state, as
demonstrated by the complex temperature dependence of the non-linear conductance, 
see Fig.~\ref{SingletSide}. 
The physical interpretation of these data is the 
following~\cite{VojtaBulla,HofstetterSchoeller,PustilnikSingletTriplet,HofstetterZarand,ZitkoSTMultidot,ZitkoSTTransport,ZarandST,Roch}: 
at high-temperatures, the two-orbitals act basically as two decoupled spin $S=1/2$, and the most 
strongly coupled spin undergoes a first screening process associated to a Kondo scale $T_{K,S=1/2}$, while
the more weakly coupled spin remains essentially free. This is witnessed in Fig.~\ref{SingletSide} by 
the formation of a standard Kondo anomaly (with maximum conductance at zero bias) for
temperatures $T>600$mK. By further cooling,
interaction between the remaining unscreened spin and the Fermi sea builds up via the
already screened $S=1/2$ moment, and leads to a second stage of screening that
fully quenches the total entropy. By adiabatic continuity to the singlet state
realized far from the transition, the conductance can be shown to drop to small
values, so that a Kondo-like dip is seen in transport associated to a second
Kondo scale $T^*$. The presence of two different energy scales is clearly shown
in Fig.~\ref{SingletSide}, while the crossover from strong singlet binding to
the Kondo-like singlet binding appears by the change from a hard gap to a softer
pseudogap in Fig.~\ref{didv}.

On the triplet side of the transition, also dispersive but less pronounced 
finite-bias excitations to the singlet state are clearly seen in
Fig.~\ref{didv}. These features are also observed, albeit more mildly, in the 
temperature-dependent zero-bias conductance by a distinctive shoulder for temperatures of 
the order of the singlet excitation threshold, see upper panel in Fig.~\ref{GT}.
Also quite striking in the experimental data on the triplet side is the ongoing 
increase with temperature of the zero bias conductance, associated to the underscreened 
Kondo effect discussed previously. We now turn to the modelization and the
microscopic analysis of this physics, giving strength to the interpretation of
the experiment in terms of the unscreening singlet-triplet quantum phase transition.

\section{Theoretical analysis of the singlet-triplet unscreening transition}
\label{Theo}

\subsection{Two-orbital Anderson model}

A quite generic model for two-electron quantum dots includes both relevant 
levels and their hybridization with the leads (see e.g. Ref.~\cite{Logan}):
\begin{equation}
H=H_{\mathrm{dot}}+H_{\mathrm{leads}}+H_{\mathrm{mix}}.  
\label{ham}
\end{equation}
The term $H_{\mathrm{dot}}$ above describes two energy levels in the quantum dot and their
mutual interactions:
\begin{equation}
H_{\mathrm{dot}}=\sum\limits_{i\sigma }\epsilon _{i}n_{i\sigma
}+\sum\limits_{i}U_{i}n_{i\uparrow }n_{i\downarrow }+U_{12}n_{1}n_{2}
-J_H\mathbf{s}_{1}\mathbf{s}_{2},  \label{hdot}
\end{equation}%
where $n_{i\sigma }=d_{i\sigma }^{\dagger }d_{i\sigma }$, $%
n_{i}=\sum_{\sigma }n_{i\sigma }$, $\mathbf{s}_{i}\mathbf{=}\sum_{\alpha
\beta }d_{i\alpha }^{\dagger }{\boldsymbol\sigma }_{\alpha \beta }d_{i\beta
}/2$, and $d_{i\sigma }^{\dagger }$ creates an electron with spin $\sigma $
on the dot level $i=$1, 2. The first term in Eq. (\ref{hdot}) describes the
single-particle energy of both levels, the second and third terms are the
intralevel and interlevel Coulomb repulsion respectively, and the last term
corresponds to the bare ferromagnetic ($J_H>0$) coupling according to the first 
Hund's rule. 
For simplicity we have omitted spin anisotropies arising from spin orbit
coupling~\cite{ParksUS,ZitkoSO,Tagliacozzo}.
The second term of Eq. (\ref{ham}) corresponds to two non-interacting leads 
[$\nu =L$ (left) or $R$ (right)]:
\begin{equation}
H_{\mathrm{leads}}=\sum_{\nu k\sigma }\epsilon _{\nu k}c_{\nu k\sigma
}^{\dagger }c_{\nu k\sigma },  \label{leads}
\end{equation}%
and the last term is the mixing (tunneling) between dot and leads 
\begin{equation}
H_{\mathrm{mix}}=\frac{1}{\sqrt{N_{k}}}\sum_{i\nu k\sigma }t_{\nu i}c_{\nu
k\sigma }^{\dagger }d_{i\sigma }+\mathrm{H.c.},  \label{hmix}
\end{equation}%
where we assume for simplicity constant tunneling amplitudes $t_{\nu i}$
near the Fermi energy, and the same density of states in each lead.

\subsection{Non-equilibrium diagrammatic calculations}
\label{diagram}

\begin{figure}
\includegraphics[scale=1.6]{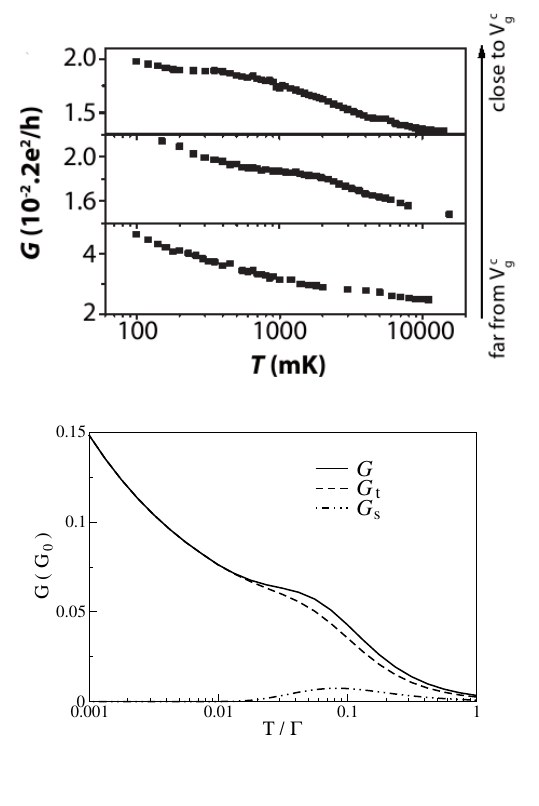}
\caption{Upper panel: experimental temperature-dependent linear conductance on the triplet 
side from far to close to the transition (bottom to top). A dispersive shoulder
associated to the underlying singlet excitation is observed. Lower panel:
theoretical calculation of $G(T)$ (full line) as performed in
Section~\ref{diagram},
showing also the relative contributions from the singlet (dash-dot-dotted line) and triplet (dashed line). 
Parameters are $D=20$, $E_t=0$, $E_s=0.1$, $\Gamma_1=1$, $\Gamma_2=1/2$.}
\label{GT}
\end{figure}

A disadvantage of the generic model (\ref{ham}) is that it has many parameters,
and shows a great deal of physical regimes, as discussed in the introduction
(see for instance Ref.~\cite{Logan} for a detailed analysis in some other 
interesting range of parameters).
The singlet-triplet quantum phase transition (QPT) takes place inside the 
$N=2$ diamond, but we have seen in Section~\ref{mechanism} that charge fluctuations 
towards the closest Coulomb diamond ($N=1$ or $N=3$) are important for
the dispersion of the magnetic excitations. Focusing on the triplet side of the 
transition from now on, we will thus retain the lowest states of 
the $N=2$ and $N=3$ configurations only.
In fact we shall show that the resulting singlet-triplet Anderson model (STAM) 
\cite{Allub} is the minimal model that describes the QPT as the gate voltage increases
from the transition point towards the triplet domain. The states and energies
retained in the STAM are the triplet state for $N=2$ composed by
\begin{equation}
|11\rangle =d_{1\uparrow }^{\dagger }d_{2\uparrow }^{\dagger }|0\rangle 
\mathrm{,} \;\; E_{t}=\epsilon _{1}+\epsilon _{2}+U_{12}-J_H/4,  \label{trip}
\end{equation}
and the states $|10\rangle $ and $|1-\!1\rangle $ obtained by applying
succesively the lowering operator $s_{1}^{-}+s_{2}^{-}$ to it, the lowest
singlet state  for $N=2$ (assuming $\epsilon _{1}<\epsilon _{2}$) 

\begin{equation}
|00\rangle =d_{1\uparrow }^{\dagger }d_{1\downarrow }^{\dagger }|0\rangle 
\mathrm{,}\;\; E_{s}=2\epsilon _{1}+U_{1};  \label{sing}
\end{equation}%
and the lowest doublet for $N=3$

\begin{equation}
|\uparrow \rangle =d_{1\uparrow }^{\dagger }d_{1\downarrow }^{\dagger
}d_{2\uparrow }^{\dagger }|0\rangle \mathrm{,}\;\; E_{d}=2\epsilon _{1}+\epsilon
_{2}+U_{1}+2U_{12},  \label{doub}
\end{equation}%
and $|\downarrow \rangle =(s_{1}^{-}+s_{2}^{-})|\uparrow \rangle $.

The problem of projecting the Hamiltonian onto this reduced Hilbert space
takes the same form as that explained in detail in Section II B of Ref.~\cite{Aligia2}, 
making an electron-hole transformation  $h_{\nu k\uparrow
}^{\dagger }=-c_{\nu k\downarrow }$, $h_{\nu k\downarrow }^{\dagger
}=-c_{\nu k\uparrow }$, $a_{\uparrow }^{\dagger }=-d_{2\downarrow }$, $%
a_{\downarrow }^{\dagger }=-d_{2\uparrow }$, $b_{\uparrow }^{\dagger
}=-d_{1\downarrow }$, $b_{\downarrow }^{\dagger }=-d_{1\uparrow }$. 

Assuming \ $t_{L1}t_{R2}=t_{L2}t_{R1}$, which corresponds to the situation of
only one screening channel discussed in Section \ref{underscreened} and defining 
$h_{\sigma}=(t_{L1} h_{L\sigma }+t_{R1}h_{R\sigma })/[t_{L1}^{2}+t_{R1}^{2}]^{1/2}$, with 
$h_{\nu \sigma }^{\dagger }=\sum_{k}h_{\nu k\sigma }^{\dagger }/\sqrt{N_{k}}$, 
the Hamiltonian takes the form of Eq. (\ref{ham}) with now

\begin{equation}
H_{\mathrm{dot}}=E_{s}|00\rangle \langle 00|+E_{t}\sum_{M}|1M\rangle \langle
1M|+E_{d}\sum_{\sigma }|\sigma \rangle \langle \sigma |,  \label{hdot2}
\end{equation}

\begin{equation}
H_{\mathrm{leads}}=-\sum_{\nu k\sigma }\epsilon _{\nu k}h_{\nu k\sigma
}^{\dagger }h_{\nu k\sigma },  \label{leads2}
\end{equation}

\begin{eqnarray}
\label{hmix2}
H_{\mathrm{mix}} &=&\Big[V_{s}\left(h_{\uparrow }^{\dagger }|\downarrow \rangle
-h_{\downarrow }^{\dagger }|\uparrow \rangle \right) \langle 00| \\
\nonumber
&+& V_{t}\left(h_{\uparrow}^{\dagger }|\downarrow \rangle 
+h_{\downarrow }^{\dagger }|\uparrow \rangle \right)\langle 10|  \\
\nonumber 
&+&\sqrt{2}V_t\left(h_{\uparrow }^{\dagger }|\uparrow \rangle \langle
11|+h_{\downarrow }^{\dagger }|\downarrow \rangle \langle 1-\!\!1|\right)\Big]+\mathrm{H.c}
,  
\end{eqnarray}%
where $V_{s}=[t_{L2}^{2}+t_{R2}^{2}]^{1/2}$ and $%
V_{t}=[(t_{L1}^{2}+t_{R1}^{2})/2]^{1/2}$.

We will show that our main experimental findings can be naturally
explained by out-of-equlibrium diagrammatic calculations using the non-crossing 
approximation (NCA) on the STAM [Eqs. (\ref{ham}), (\ref{hdot2}), (\ref{leads2}) 
and (\ref{hmix2})], following our recent theoretical work~\cite{Aligia1,Aligia2}
(some aspects of the equilibrium situation are also reported in NRG
simulations in Refs.~\cite{HofstetterSchoeller,Pruschke,ZitkoSTMultidot,ZitkoSTTransport} and in Sec.~\ref{NRG} 
below, as well as in the related study by Logan {\it et al.}~\cite{Logan}). 
Let us now review our theoretical results, skipping the mathematical details, and try 
to connect them to the experimental discussion of section~\ref{exp}.

\begin{figure} 
\includegraphics[scale=0.23]{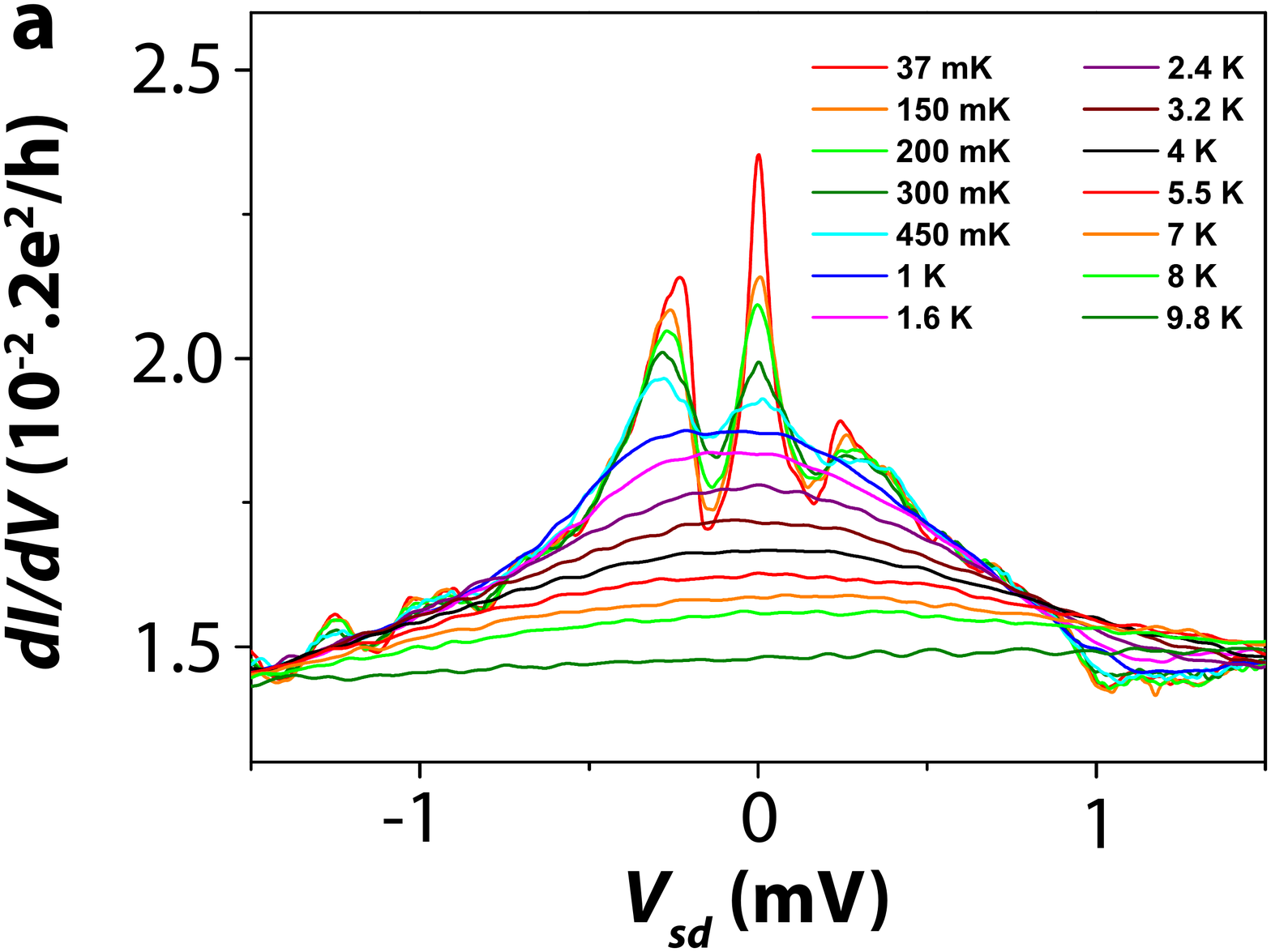}
\includegraphics[scale=0.23]{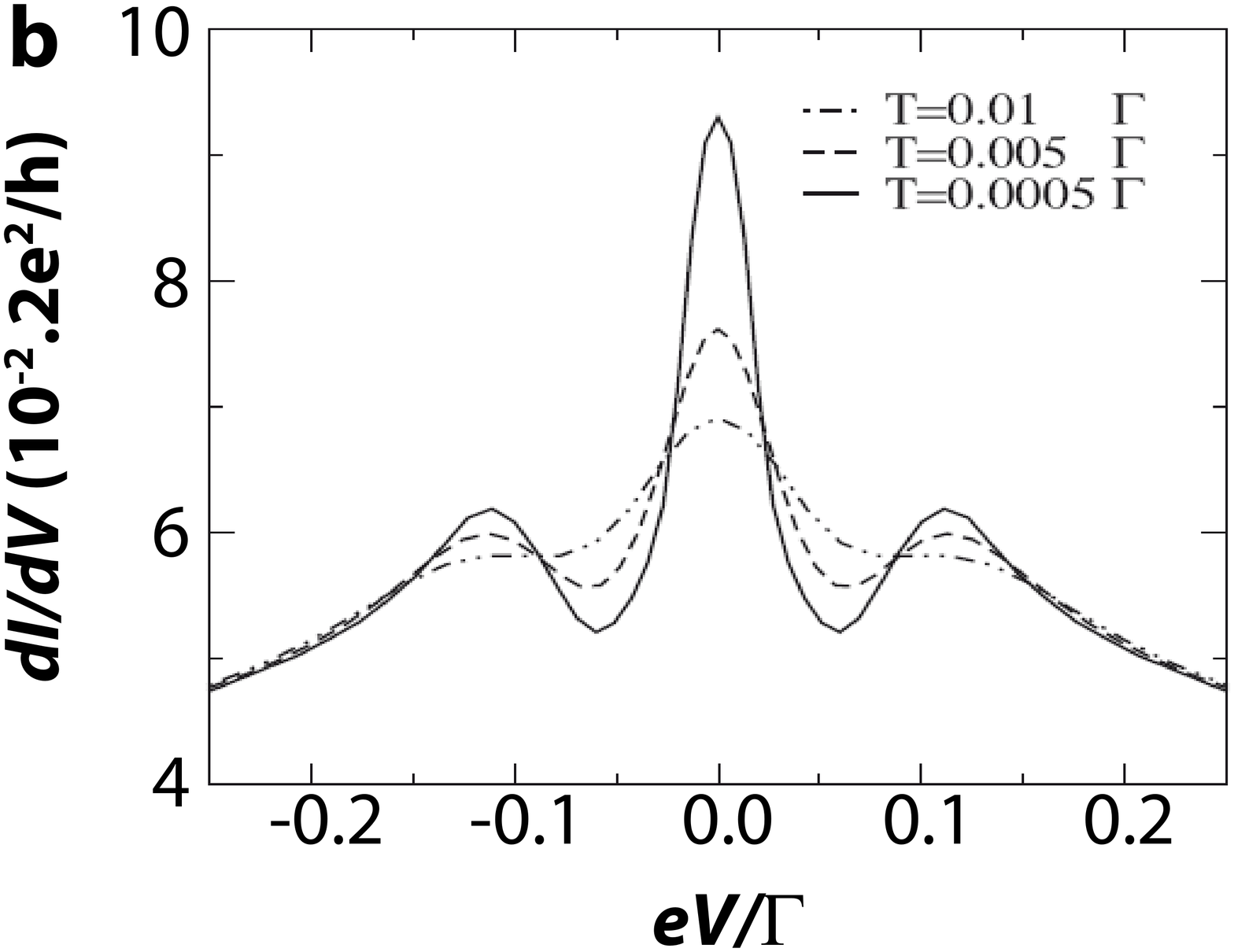}
\caption{Differential conductance $dI/dV$ as a function of voltage on the triplet side of the
singlet-triplet transition for different temperatures. The experimental data \cite{Roch} is
shown on panel a, and the calculations \cite{Aligia1} on panel b are
made for the same parameter as Fig.~\ref{GT}.}
\label{threepeaks}
\end{figure}

Recent theoretical work by two of us~\cite{Aligia1,Aligia2} has confirmed
one striking experimental observation pertaining to the finite bias conductance,
namely the emergence, as a function of temperature, of a three peaks structure on
the triplet side of the QPT, see Fig.~\ref{threepeaks} comparing the data and
the numerics (note that the data correspond here to a gate voltage taken on the 
triplet side of the transition in the lower panel of Fig.~\ref{diamond}).
In this theoretical calculations, as in those discussed below, a symmetric coupling of 
the dot with both left and right leads has been assumed for simplicity.
Allowing an asymmetric coupling manages to improve the comparision with experiment 
(see Fig. 11 of Ref.~\cite{Aligia2}). 
Studying now the temperature-dependent linear conductance, a shoulder-like
feature is obtained from the calculation (lower panel of Fig.~\ref{GT}), quite similar to
the experimental observation of Fig.~\ref{GT} (upper panel). This is explained by the
admixture of the triplet and singlet excitations at high temperatures. For
lower temperature than the triplet-singlet splitting, the conductance continues
to increase, due to the triplet Kondo effect. We note that underscreening is 
not quantitatively captured by the NCA, which misses the logarithmic saturation 
towards the zero temperature conductance (not shown).

These previous calculations have been performed for equal hybridization strength
$V_{s}=$ $V_{t}$, and in this case the distance to the QPT is simply controled by the parameter 
$\delta=E_{t}-E_{s}$. Furthermore, at the quantum critical point ($\delta =0$), the model 
is exactly solvable \cite{Allub,Aligia2} allowing to test the diagrammatic calculations
\cite{Aligia2}. However, experimentally the {\it bare} splitting $\delta $ is approximately 
constant as the gate voltage $V_{g}$ is varied, while the renormalized
singlet-triplet splitting is gate-dependent due to the orbital asymmetry, so
that the triplet is lowered in energy with increasing $V_{g}$. In our
calculations for positive $\delta$ the QPT transition
can take place as a function of increasing $V_{g}$ if $V_{t}>V_{s}$
(corresponding to $t_{L1}>\sqrt{2}t_{L2}$) as explained in Section \ref{mechanism}. 
Actually, the STAM for $\delta >0$ and $V_{t}>V_{s}$ has also been proposed for Tm
impurities and studied with NRG fifteen years ago \cite{Allub}. 
As an example, defining  
\begin{eqnarray}
\Gamma _{i} &=&2\pi \sum_{k\nu }|t_{\nu i}|^{2}\delta (\omega +\epsilon
_{\nu k}),   \\
\Delta  &=&E_{d}-(E_{s}+3E_{t})/4,  \label{def}
\end{eqnarray}%
with the Fermi level as the origin of energies, for $\Gamma _{2}=\Gamma
_{1}/4$, $\delta =0.345\Gamma _{1}$ and a band width $D=100\Gamma _{1}$, the
quantum critical point (QCP) determined by NRG \cite{Allub} is at  $\Delta =$
$\Gamma _{1}$. For smaller (larger) values of $\Delta $, the ground state of
the system is a doublet (singlet) and the system is in what was denoted as
the triplet (singlet) side of the transition.

To extend these calculation to the non-equilibrium situation and to compare
with the experimental non-linear conductance (upper panel of Fig. \ref{didv}), we have used 
the NCA to calculate the differential conductance $G(V)$ for a new set of parameters with 
$\delta >0$ and $V_{t}>V_{s}$, in which the transition is driven by gate
voltage, thus shifting the singlet-triplet splitting as in the experiment. 
The results are shown in the lower panel of Fig. \ref{didv}. 
The quantum critical point is at $E_{d}=E_{d}^{c}\simeq 4.3$. For $E_{d}\leq
E_{d}^{c}$, $G(V)$ shows a Kondo peak at $V=0$ due to the partial screening
of the spin $S=1$ localized states. As $E_{d}^{c}-E_{d}$ increases into the triplet
region a mild peak appears at finite bias and displaces to larger values of $V$, 
which corresponds to finite energy singlet excitations \cite{Aligia1,Aligia2}. 
For  $E_{d}>E_{d}^{c}$, the ground state is a singlet and the conductivity
drops at $V=0$ to a value which depends on the occupation and is
determined at $T=0$ by the Friedel-Luttinger sum rule \cite{Logan,Aligia2}.
As $E_{d}-E_{d}^{c}$ increases further in the singlet region, a sharp Kondo-enhanced 
\cite{Paaske} finite-bias conductance peak develops due to excitations of triplet character 
\cite{Aligia2}. This feature becomes sharper and displaces to larger $V$ far from the
critical point, in agreement with the experimental data of Fig.~\ref{didv}
(upper panel). 
Overall, the qualitative agreement between theory and experiment is good. There is a
discrepancy in the evolution of the value of $G(V)$ at its maximum, which
increases sharply as the experimental system is driven deep inside the singlet
side of the transition, while it decreases slowly within our NCA calculation. Because
our truncated STAM Hamiltonian considers only the $N=2$ and $N=3$ states, and
neglects the charge $N=1$ configuration, valence fluctuations are artificially 
suppressed in the singlet phase. Hence, the Kondo scale associated to the enhancement 
of the out-of-equilibrium cotunneling line is underestimated in the calculation, so
that the finite-bias peak cannot sharpen by approaching the diamond edge.
Inclusion of the charge $N=1$ states is infact be required for a better 
modelization of the data, as we will now see.

\subsection{Numerical Renormalization Group analysis}
\label{NRG}

We are coming back now to the complete model~(\ref{ham}) which contains
all possible charge states, from zero to four electrons, in the two-orbital 
Anderson model. The goal here is to present a more global view of the transport 
properties of such quantum dot, extending thus our previous theoretical analysis 
to the neighboring diamonds, as well as giving a confirmation of the results
from the diagrammatic method within the singlet-triplet sector of the two-electron 
configuration. The technique which we use from now on is the Numerical Renormalization 
Group (NRG), see Refs.~\cite{Wilson,BullaRMP} for a review. Interesting studies 
of the properties of model~(\ref{ham}) within the NRG, as well as from perturbative arguments, appeared in 
Refs.~\cite{HofstetterSchoeller,Pruschke,Logan,ZitkoSTMultidot,ZitkoSTTransport} but 
within a parameter range that does not fully correspond to the present experimental conditions.
We thus wish here to present an investigation of transport in the situation of asymmetrically
coupled orbitals, leading to the gate-controlled singlet-triplet
transition~\cite{ZitkoSTMultidot,Logan}, which was not completed in full detail in these 
previous works.

It is important to make a few words on methodology here. At present, a fully
non-equilibrium generalization of the NRG is still lacking, despite recent
advances~\cite{Anders}, so that we will consider a near-equilibrium situation.
This approximation is relatively well-obeyed, owing to the non-unitary
conductance of the spin $S=1/2$ Kondo ridge (see Fig.~\ref{diamond}), that
result from a generic asymmetric coupling of the molecule to the source and
drain.
We also assume that a single channel couples to the impurity states, which
seems also the case in the experiment, otherwise the underscreened Kondo effect
would be spoiled at low temperatures. This situation can be achieved for instance by
considering the following constraint among the tunneling matrix element: 
$t_{Li}=\cos(\phi) t_i$ and $t_{Ri}=\sin(\phi) t_i$, with $\phi$ an arbitrary
angle that encodes the maximum value $G_0$ of the conductance in the device. 
Indeed, from the hybridization to each lead $\Gamma_L=\pi \rho_0
(t_{L1}^2+t_{L2}^2)$ and $\Gamma_R=\pi \rho_0
(t_{R1}^2+t_{R2}^2)$ with $\rho_0$ the density of states in the leads, one
gets $G_0=(2e^2/h) 4\Gamma_L\Gamma_R/(\Gamma_L+\Gamma_R)^2=(2e^2/h)
\sin^2(2\phi)$.
One can then perform a linear combination of the left and right lead electrons
to end up with an effective single-channel two-impurity Anderson model with
tunneling elements $t_1$ and $t_2$ onto each
orbital~\cite{HofstetterSchoeller,ZitkoSTTransport,Logan}.
One can compute then the total spectral density (or T-matrix) within 
the standard NRG at equilibrium~\cite{HofstetterSchoeller,Logan}
\begin{equation}
\label{Tmatrix}
\rho(\w)=-\frac{1}{\pi}\sum_{i,j=1,2}
\frac{\Gamma_{ij}}{\Gamma_{11}+\Gamma_{22}} \,
\mathcal{I}m G_{ij}(\w)
\end{equation}
where the hybridization matrix reads $\Gamma_{ij}=\pi\rho_0 t_i t_j$, and
the retarded Green's function $G_{ij}=\big<\big<d^{\phantom{\dagger}}_i ;d^\dagger_j\big>\big>$
is computed on the real frequency axis.
This spectral density $\rho(\w)$ requires very high resolution data in order
to describe reliably both the diamond edges (linked to the Hubbard satellites) and
the fixed-charge cotunneling features within the Coulomb diamonds. In order to obtain 
the latter, the extensive use of an optimized broadening method, the so-called ``b-trick''
\cite{Freyn}, turns out to be crucial (technical detail will appear 
elsewhere~\cite{Freyn2}).
One finally uses the Landauer-Meir-Wingreen formula~\cite{MeirWingreen,Logan} in order
to compute the finite-bias conductance:
\begin{eqnarray}
\label{Landauer}
G(T,V_{sd}) & = & G_0 \int_{-\infty}^{+\infty} \!\!\!\!\!\!\! d\w 
\; \pi (\Gamma_{11}+\Gamma_{22})\rho(\w) \\
\nonumber
&& \hspace{-0.5cm} \times
\left(-\frac{1}{2}\right)
\left[\frac{\partial n}{\partial \w}(\w+V_{sd}/2)+
\frac{\partial n}{\partial \w}(\w-V_{sd}/2)\right]
\end{eqnarray}
where $n(\w)=1/(\exp(\w/T)+1)$ is the Fermi function.

\begin{figure}
\includegraphics[scale=1.1]{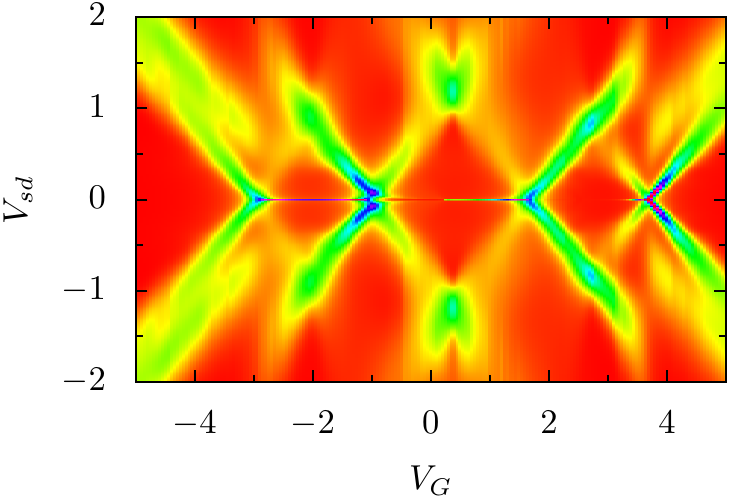}
\caption{Conductance obtained by the NRG for the two-orbital Anderson
model~(\ref{ham}) at zero temperature, with parameters $U_{11}=U_{22}=0.5U_{12}=1$, 
$J_H=3$, $t_1=0.25$, $t_2=0.125$, $\epsilon_2-\epsilon_1=0.742$ (all in units of the 
half-bandwidth of the leads), giving rise to a complex Coulomb diamonds diagram.
One readily identifies the usual charging pattern as a function of gate voltage 
$V_g$ and source-drain bias $V_{sd}$, showing that the dot can accomodate from zero
electron (for the most negative $V_g$) up to four electrons. However, more
complex structures are also noted, such as sequencial tunneling lines that run
parallel to the $N=0$ and $N=4$ diamond edges as well as Kondo features within the 
charge $N=1,2,3$ diamonds. A close-up within the middle diamond containing two 
electrons is given in Fig.~\ref{NRG2}.}
\label{NRG1}
\end{figure}

Because of the near-equilibrium approximation for the T-matrix, we cannot pretend to 
a full quantitative agreement with the experimental data, as was however achieved for
the Kondo related features at zero bias~\cite{RochUS}.
However, the precise determination of the many parameters in the two-orbital
Anderson model~(\ref{ham}) remains a very challenging task, so that we can only hope
to capture the qualitative details of the experiment. An important
point that we wish also to stress is that the observed experimental features
possess remarkable universal features, so that the exact determination of the
microscopic couplings in the model is not too important. We have already
emphasized here and in Sec.~\ref{underscreened} that spin $S=1/2$ and $S=1$ Kondo anomalies
provide universal scaling function for the temperature dependent conductance,
that are faithfully captured by the NRG calculations~\cite{RochUS}. We will
further demonstrate here that the gate-dependence of the singlet to triplet
cotunneling excitations is also a very robust property in the model~(\ref{ham}),
as soon at the tunnel couplings $t_{\nu 1}$ and $t_{\nu 2}$ between the leads
($\nu=L,R$) and each of the orbitals  are made asymmetric, in agreement with
general experimental observations of this phenomenon\cite{Holm,Hauptmann,Roch,Osorio}.

\begin{figure}
\includegraphics[scale=1.1]{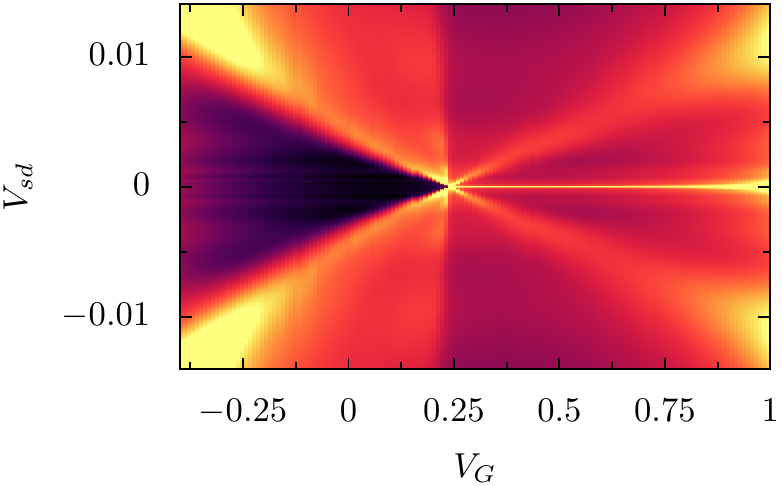}
\caption{Singlet-triplet unscreening quantum phase transition obtained from the
NRG calculations at zero temperature (with the same parameters as in Fig.~\ref{NRG1}). Finite bias cotunneling
features, with sharp Kondo-like enhancement, are obtained thanks to optimized
broadening methods~\cite{Freyn}. The zero-bias Kondo anomaly for gate voltages
$V_g>0.25$ is associated to the underscreening of the triplet state, while a
sharp gap occurs on the singlet side of the transition for $V_g<0.25$.}
\label{NRG2}
\end{figure}

Let us first discuss the zero-temperature global transport ``phase diagram'' of the two impurity
Anderson model, see Fig.~\ref{NRG1}. Several Coulomb diamonds, showing a charging pattern
from zero to four electrons as a function of gate voltage $V_g$, are obtained as 
expected. Sequential tunneling lines~\cite{Hanson}, running parallel to the
edges of the diamonds with $N=1$ and $N=4$ electrons are also seen, and can be attributed to
the level splitting $|\epsilon_2-\epsilon_1|$ between the two orbital states.
Although these latter features could be anticipated from general grounds, their
observation within an NRG calculation was not reported to our knowledge in
previous studies~\cite{HofstetterSchoeller,ZitkoSTTransport,Pruschke}. One can also barely guess
that this line moves within the $N=1$ diamond as a smoother cotunneling line, an
effect that is also seen in the experimental data in Fig.~\ref{diamond}.
Several Kondo anomalies are
also observed at low bias: i) in the charge $N=1$ diamond, the well-studied $S=1/2$ Kondo
effect occurs, with full screening of the impurity spin; ii) in the charge $N=2$
diamond, a singlet-triplet transition can be already guessed, see also the
close-up in Fig.~\ref{NRG2} and further discussion below; iii) in the charge $N=3$ diamond, 
a complex Kondo state occurs, as previously examined in Refs.~\cite{PustilnikBorda,Logan}.
In this last case, a spin $S=1/2$ state is realized by totally filling the lowest orbital
with two electrons and adding an unpaired electron on the upper orbital. Nevertheless the 
resulting Kondo screening process turns out to be subtle at the boundary between the $N=2$ and $N=3$
diamonds. This is because full screening of the $S=1/2$ moment provides a zero
entropy ground state within the $N=3$ diamond, while the $S=1$ underscreened Kondo
effect within the $N=2$ diamond has a remanent $\log(2)$ entropy, leading to a
second quantum critical point precisely at the edge between both diamonds, besides the 
critical point located within the $N=2$ diamond, that we discuss now.

We finally focus the discussion on the charge $N=2$ configuration, for which the fine 
structures are given in Fig.~\ref{NRG2}.
The singlet-triplet unscreening transition is first evidenced by the
obvious dichotomy in the low bias transport properties: a low conductance state in 
the singlet sector (gate voltage $V_g<0.25)$  is replaced by a highly conducting 
spin $S=1$ Kondo anomaly in the triplet sector. Concomitantly, a clear crossing of
excited states occurs precisely at the quantum critical point. These cotunneling
lines show sharp Kondo-like enhancement, as discussed previously with respect to
both the experimental data (Sec.~\ref{exp}) and the diagrammatic calculations
(Sec.~\ref{diagram}). Clearly, the inclusion of the charge $N=1$ state in
the NRG simulations allows us to recover the correct enhancement of the
satellites on the singlet side as the gate voltage is decreased towards the
diamond edge. This property is also crucial to obtain theoretically the bending of the cotunneling 
line near the edge, that is also clearly seen in the experiment (see
Fig.~\ref{gatetuned}).
On a technical level, we emphasize that such numerical plot requires to perform the broadening of the NRG
raw data with the so-called b-trick~\cite{Freyn}, otherwise the cotunneling
lines are barely resolved. On a more physical ground, we have checked by
varrying the microscopic parameters of the model that the gate-induced
singlet-triplet transition remains robust, as long as the bare singlet-triplet
gap does not overcome the energy gain associated to the asymmetric tunneling
probabilities (in agreement with the general discussion in Sec.~\ref{mechanism}).
This observation highlights the singlet-triplet transition as an almost {\it generic}
feature of two-electron quantum dots.

\section{Review of various scenarios and possible instabilities in a two-electron quantum dot}
\label{instabilities}

We would like to finish by a general discussion of the possible mechanisms that
can drive instabilities in quantum dots with two electrons, in cases where only unique 
singlet and triplet states are involved. Several scenarios can be distinguished,
whether a single or two screening channels are active, and depending if a Zeeman
effect splits the triplet levels or not. The additional role of magnetic
anisotropies due to spin-orbit coupling will also be considered.

\subsection{Singlet-triplet ``transition'': single channel case}
\begin{figure*}[ht]
\centerline{\includegraphics[scale=0.45]{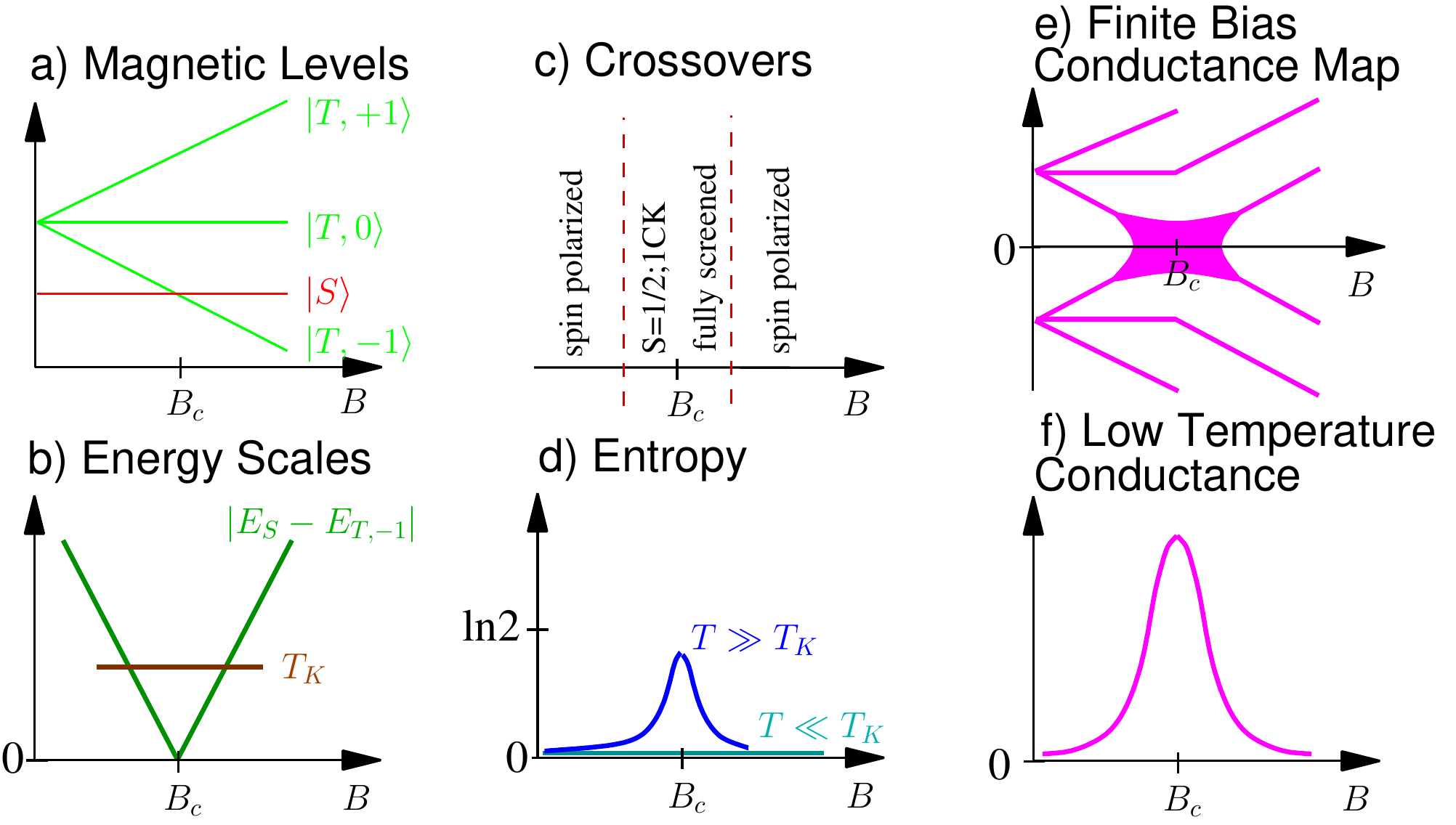}}
\caption{Singlet-triplet ``transition'' in the single-channel case: Zeeman induced Kondo 
effect by crossing of singlet and lowest triplet state.} 
\label{ST_transition_1CK}
\end{figure*}

We will start by discussing the situation of the so-called singlet/triplet ``transition'', 
which is rather a magnetic-field induced Kondo effect due to the extra
degeneracy achieved at the crossing of singlet and triplet states. Several 
proposals for this effect have been made, and we present first the simplest
one, which involves a Zeeman-split triplet, in the single channel case 
\cite{Nygard,PustilnikAvishai}. The role of the magnetic field here is to create an effective 
doublet by bringing the lowest triplet state in degeneracy with the singlet state 
(see panel a in Fig.~\ref{ST_transition_1CK}).
This effective doublet is then screened by the conduction electrons in a standard
Kondo process, giving rise to a single-channel Kondo scale $T_K$, while
deviations from the crossing point (a role played by the singlet-triplet splitting 
$|E_S-E_{T,-1}|$) act as an effective magnetic field (see panel b in 
Fig.~\ref{ST_transition_1CK}). As this energy scale becomes larger than $T_K$,
the Kondo resonance is killed, and the linear conductance decreases (see panel f
in Fig.~\ref{ST_transition_1CK}).
Clearly, there is no quantum phase transition in this problem, but rather
smooth crossovers when the two relevant energy scales match together (see panel c in
Fig.~\ref{ST_transition_1CK}). This is because the ground state remains a spin singlet 
(zero entropy state) throughout the whole phase diagram. This of course is true
including the degeneracy point, because Kondo screening quenches the entropy of the 
doublet (see panel d in Fig.~\ref{ST_transition_1CK}).

We finally note that this effect has been studied experimentally in carbon
nanotube quantum dots~\cite{Nygard}, and should be observable typically in
molecular devices as well, due to the higher g-factor compared to semiconducting
systems. Clearly our device \cite{Roch}, taken on the singlet side and with a
strong magnetic field, satisfies most of the conditions for this scenario (see the 
similarity of panel b in Fig.~\ref{zeeman} and panel e in Fig.~\ref{ST_transition_1CK}).
However, the Kondo enhanced crossing of the singlet and lowest triplet predicted
above \cite{PustilnikAvishai} is clearly missing in our experiment, because the two 
orbitals seem to be quite asymmetrically coupled to the leads, so that the resulting 
Kondo temperature is extremely low.

\subsection{Singlet-triplet ``transition'': two-channel case}

\begin{figure*}[ht]
\centerline{\includegraphics[scale=0.45]{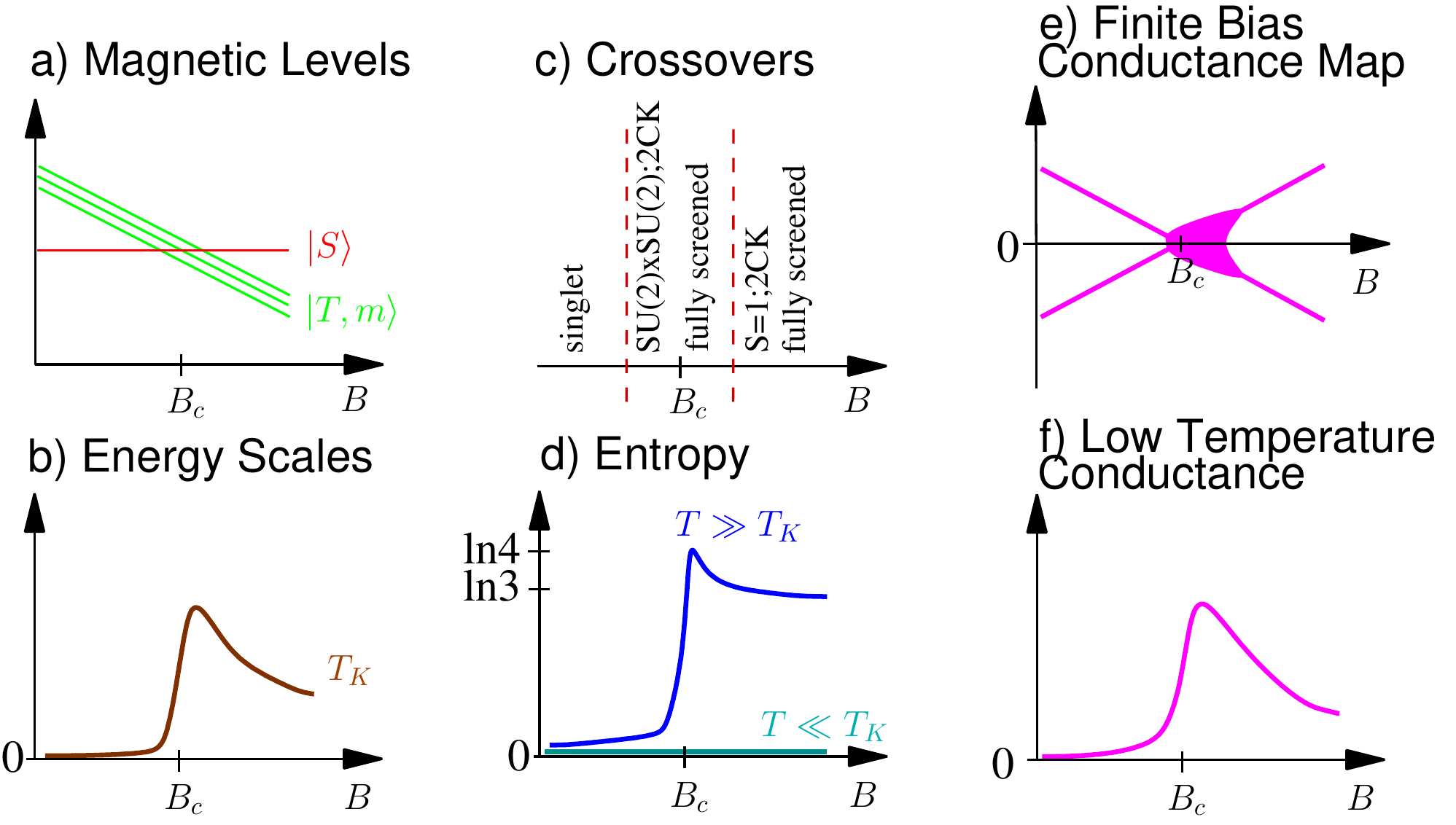}}
\caption{Singlet-triplet ``transition'' in the two-channel case: 
magnetic-field induced Kondo effect by crossing of singlet and
threefold degenerate triplet states.}
\label{ST_transition_2CK}
\end{figure*}

An alternative scenario for the singlet-triplet transition, which involves
different degrees of freedom (namely a singlet and a {\it three-fold degenerate}
triplet) has also been 
considered~\cite{STKondo_Eto,STKondo_Pustilnik1,STKondo_Pustilnik2},
following experiments in vertical quantum dots~\cite{Sasaki}.
In that case, orbital quantum numbers are conserved during the tunneling processes,
so that two screening channel are active. The control parameter here is
a perpendicular magnetic field, which by orbital effects tunes the
singlet-triplet gap (see panel a in Fig.~\ref{ST_transition_2CK}).
Due to the presence of {\it two} screening channels, the triplet phase is fully 
screened (this remains true at the degeneracy point, where a SU(2)$\times$SU(2) 
state is generated), so that the ground state entropy remains zero at all values 
of the magnetic field, and again crossovers rather than quantum phase transition 
are expected (see panels c and d in Fig.~\ref{ST_transition_2CK}). 
Because the triplet phase involves a Kondo process, the Kondo temperature is enhanced
compared to the singlet phase (see panel b in Fig.~\ref{ST_transition_2CK}). 
The low-temperature conductance thus shows an asymmetric maximum as a function 
of magnetic field (see panel f in Fig.~\ref{ST_transition_2CK}), which
distinguishes this case from the single channel situation discussed previously. 
We note that magnetic transitions appear also differently (compare panel e in
Fig.~\ref{ST_transition_1CK} and Fig.~\ref{ST_transition_2CK}).

\subsection{Singlet-triplet unscreening quantum phase transition}
\begin{figure*}[ht]
\centerline{\includegraphics[scale=0.45]{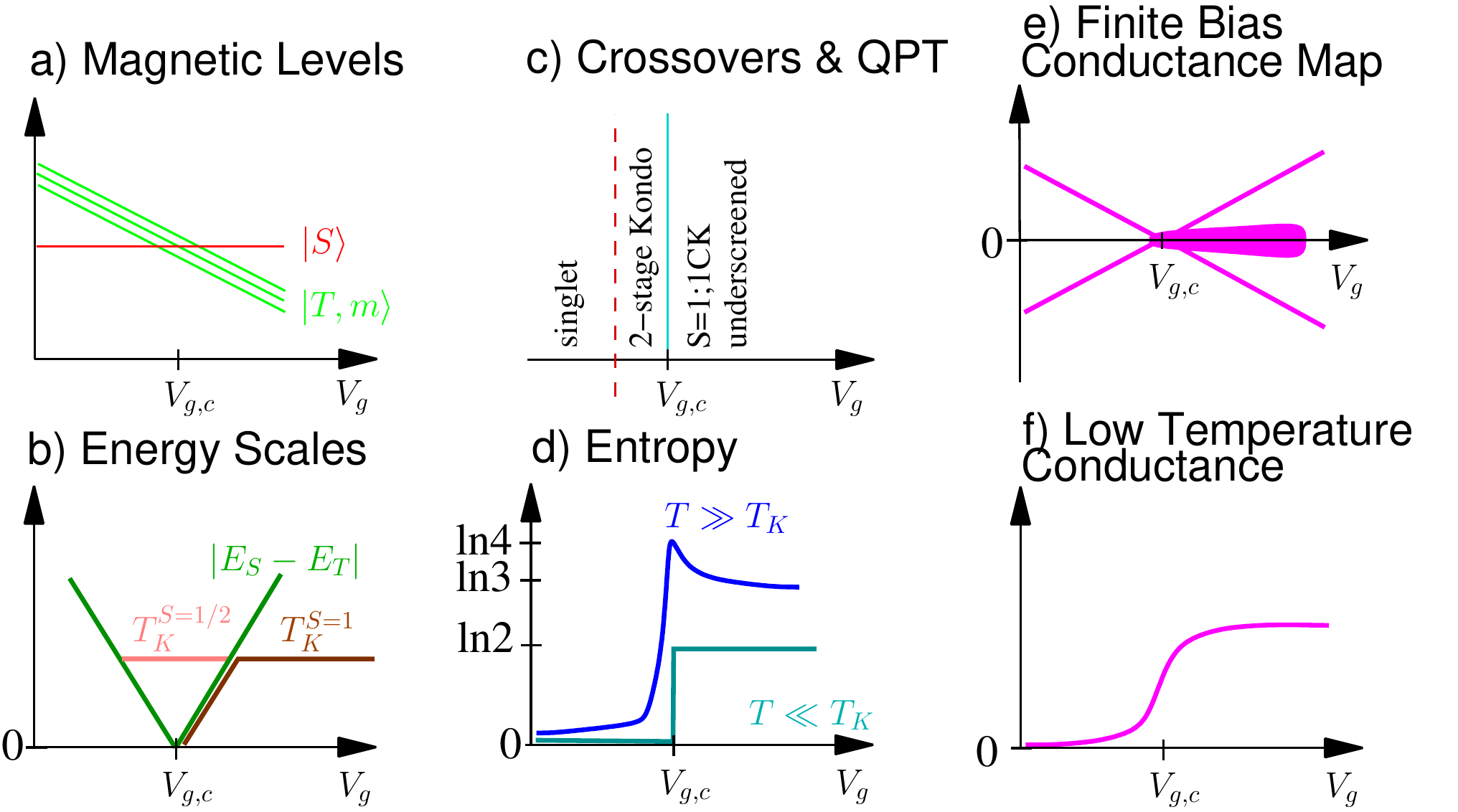}}
\caption{Singlet-triplet quantum phase transition: gate-induced crossing of singlet 
and three-fold degenerate triplet with a single screening channel}
\label{ST_QPT}
\end{figure*}

The situation that we have examined in our experiment~\cite{Roch} is markedly
different from the two scenarios discussed before: the gate-tuned singlet-triplet 
transition involves the crossing of a singlet with a {\it three-fold} degenerate triplet 
(see panel a in Fig.~\ref{ST_QPT}) in the presence of a {\it single} screening channel.
The physics is then more exotic on several accounts. First, the triplet
phase displays an underscreened Kondo effect~\cite{NozieresBlandin}, so
that a zero-bias peak coexists with the finite bias singlet excitations
(see lower panel in Fig.~\ref{diamond} and panel e in Fig.~\ref{ST_QPT}).
We have later on confirmed in detail this underscreened state,
see discusion in Sec.~\ref{underscreened} and Ref.~\cite{RochUS}.
Because of the partial screening of the $S=1$ spin, the entropy of the triplet 
phase saturates to $\log(2)$ at low temperature, so that there is an entropy 
change to zero at the singlet-triplet crossing (see panel d in Fig.~\ref{ST_QPT}). 
This corresponds to a zero temperature singlet-triplet quantum phase transition
\cite{Allub,VojtaBulla,HofstetterSchoeller,PustilnikSingletTriplet,HofstetterZarand,ZitkoSTMultidot,ZitkoSTTransport,ZarandST},
which is manifested by a step like behavior of the linear conductance as a function of gate 
voltage (see panel f in Fig.~\ref{ST_QPT}). The situation on the singlet side
close to the quantum critical point is also intriguing, because singlet and
triplet states are released into two independent spin $S=1/2$, that are screened
in a two-stage Kondo process.
The resulting energy scales and the associated scenario are given on panel 
b and c in Fig.~\ref{ST_QPT} respectively. One can also understand the
singlet-triplet QPT by an unscreening transition, where the underscreened
Kondo phase corresponds to the ferromagnetic side of an {\it effective}
$S=1/2$ Kondo model, while the singlet binding is associated to a Kondo screening
process (antiferromagnetic side in the effective model)
\cite{HofstetterSchoeller,ZitkoSTMultidot}, see the schematic flow diagram in Fig.~\ref{Flow}.

\subsection{Zeeman splitting of the underscreened Kondo anomaly}
\begin{figure*}
\centerline{\includegraphics[scale=0.45]{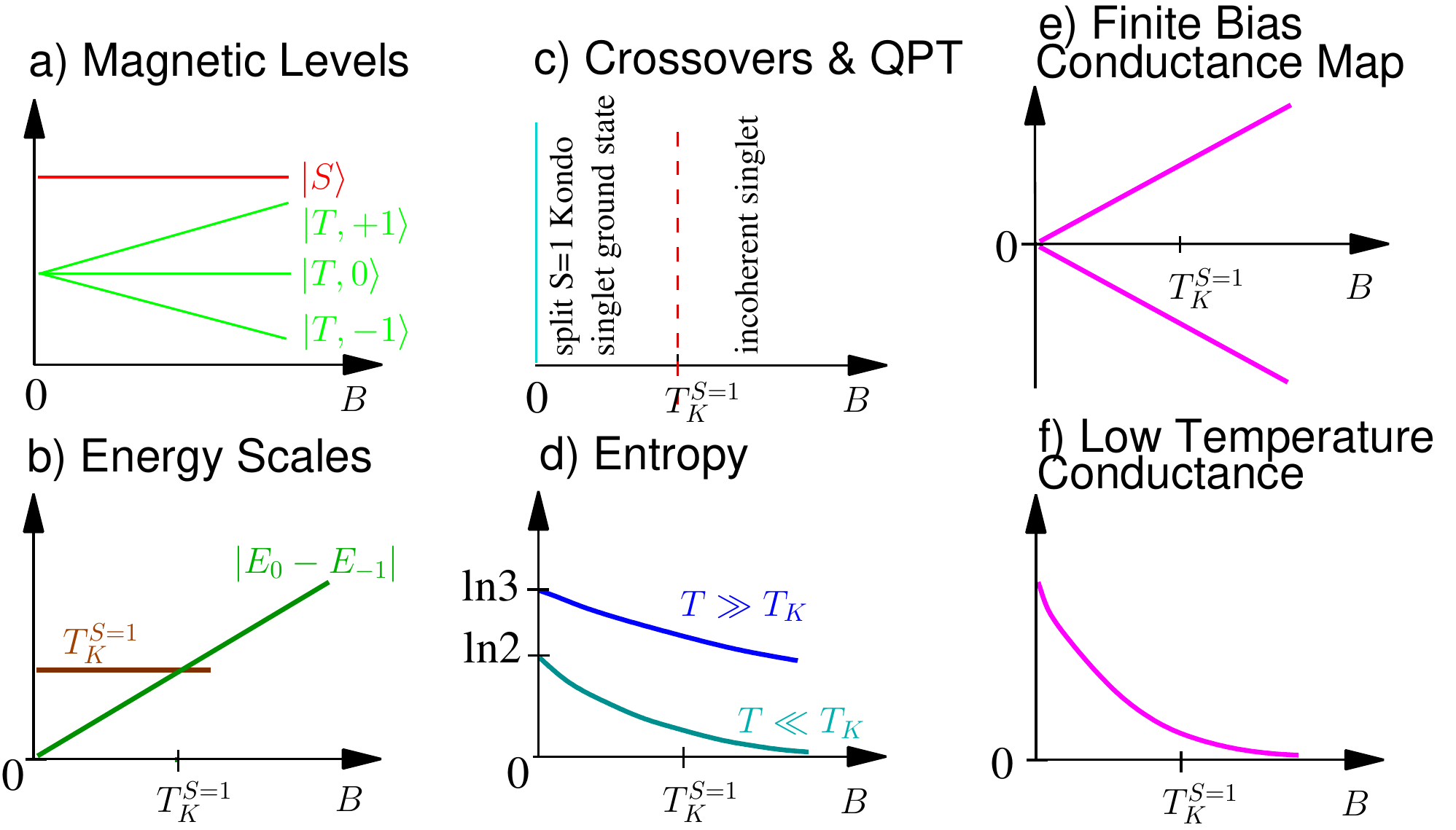}}
\caption{Magnetic field splitting of the underscreened $S=1$ Kondo state.}
\label{USK_Zeeman}
\end{figure*}

We comment here on the effect of a magnetic field onto the underscreened
$S=1$ Kondo state. The standard (fully screened) $S=1/2$ Kondo state, governed
by Fermi liquid theory, is known to present a threshold in magnetic field
(of the order of the Kondo temperature) at which the finite bias conductance
exhibits a splitting~\cite{GrobisReview}. 
In contrast, we have found both experimentally and theoretically~\cite{RochUS} that 
the underscreened state has a dramatic sensitivity to magnetic field, due to the 
finite $\log(2)$ entropy in the ground state, see Fig.~\ref{USK_Zeeman}.
In that case, the non-linear conductance splits (at zero temperature) for
arbitrarily small values of the magnetic field.

\subsection{Magnetic anisotropies driven quantum phase transition}
\begin{figure*}
\centerline{\includegraphics[scale=0.45]{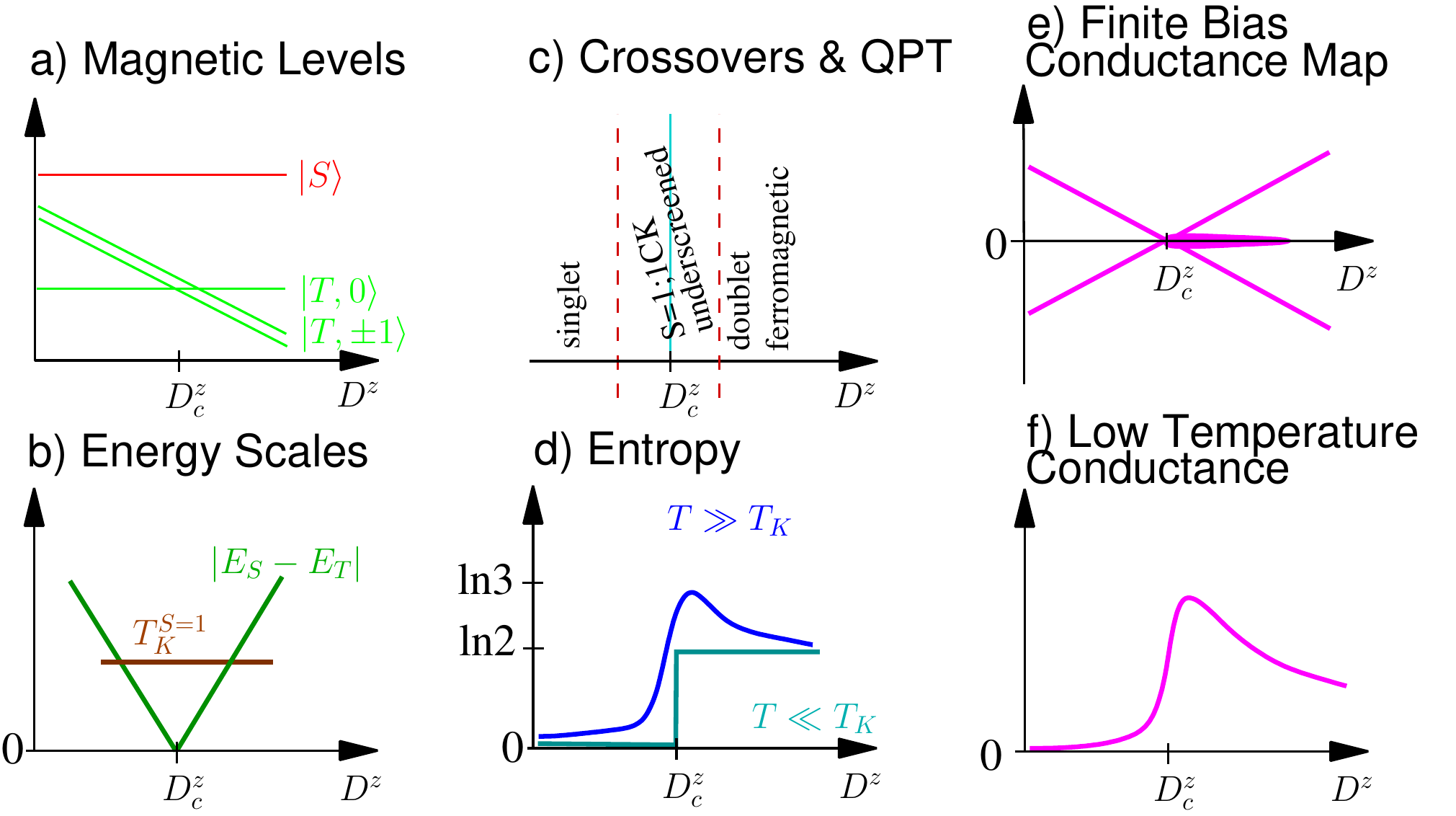}}
\caption{Spin-orbit quantum phase transition: underscreened Kondo anomaly in
the presence of a tunable magnetic anisotropy $D^z$.}
\label{SpinOrbit_QPT}
\end{figure*}

\begin{figure}
\includegraphics[scale=0.25]{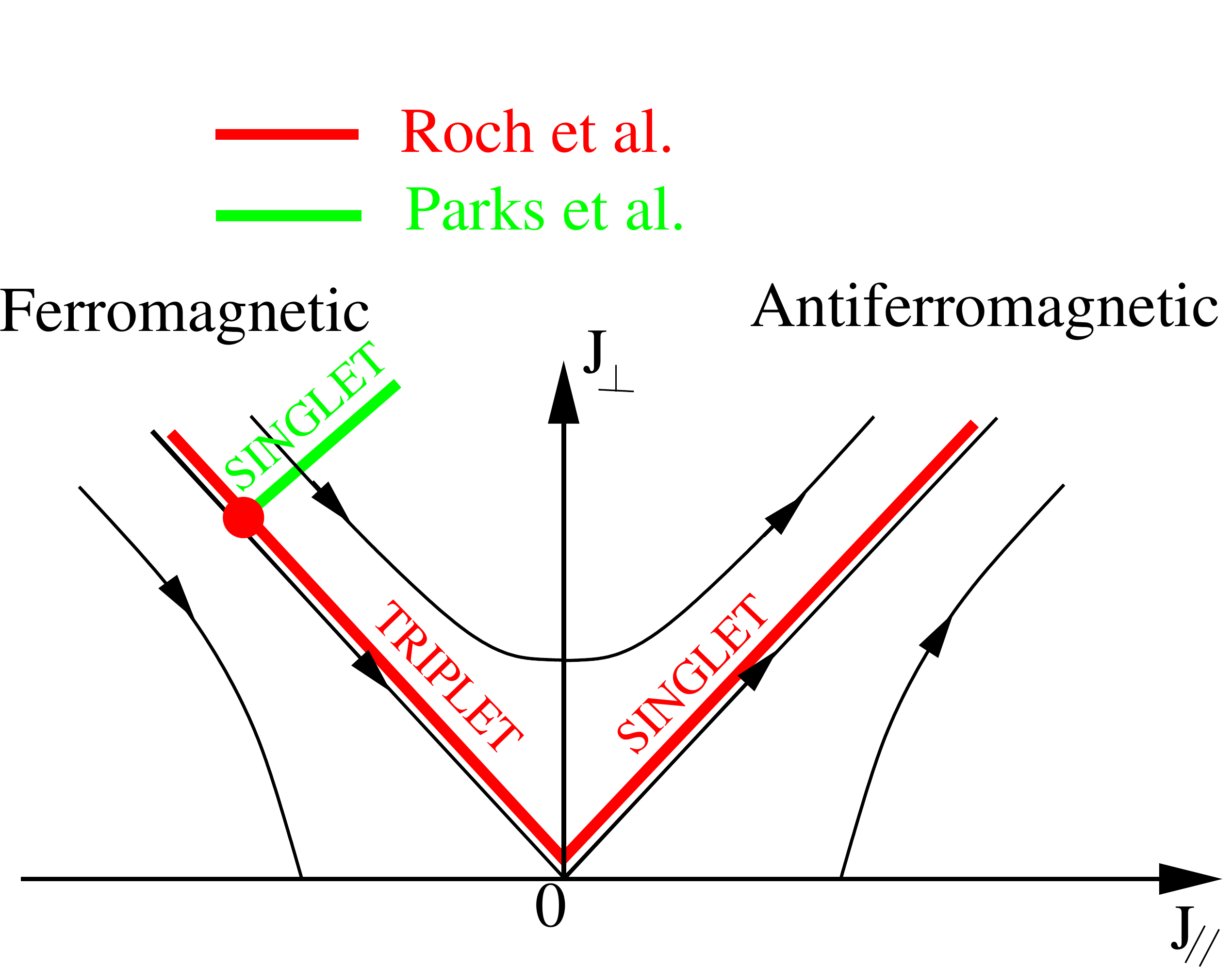}
\caption{Flow diagram associated to our experiment Refs.~\cite{RochUS,Roch}
and the experiment by Parks {\it et al.} \cite{ParksUS}.}
\label{Flow}
\end{figure}

We end up by a discussion of possible relevant perturbations to the
physics discussed above, that are related to spin orbit interactions.
Recent experiments \cite{DeFranceschi,Jespersen,ParksUS} and theoretical 
works \cite{ZitkoSO,Galpin,Tagliacozzo,ParksUS,Cornaglia,Andergassen} have demonstrated
that magnetic anisotropies can strongly affect the nature of the magnetic 
states in a two-electron quantum dot. Let us consider the addition
of a magnetic anisotropy term $D^z (S_z)^2$, which splits the
triplet into unpolarized $\big|T,0\big>$ and polarized $\big|T,\pm1\big>$ 
states (see panel a in Fig.~\ref{SpinOrbit_QPT}). 
The experiment by Parks {\it et al.} \cite{ParksUS} corresponds precisely 
to the situation of $D^z>0$, where a spin $S=1$ Kondo anomaly is 
progressively destroyed by lowering in energy the unpolarized state, so
that the conductance is suppressed (see panel f in Fig.~\ref{SpinOrbit_QPT}).
This is experimentally realized by tuning the magnetic anisotropy with a 
stretching of their molecule, and this realizes the schematic flow diagram 
in Fig.~\ref{Flow}. In contrast to our experiment where the underscreened Kondo 
was associated to a whole phase of the flow diagram, the Parks experiment
realizes the underscreened Kondo as a quantum critical point associated
to a Kosterlitz-Thouless QPT. However, the doublet phase corresponding to 
$D^z<0$ was not accessible in this experiment, so that the QPT was not
fully examined. As was noted previously~\cite{ZitkoSO,Tagliacozzo}, the
magnetic anisotropy term is less relevant than the Zeeman splitting discussed
in the previous paragraph and in our work~\cite{RochUS}, leading to a threshold
behavior at $D^z\simeq T_K^{S=1}$ for the splitting of the non-linear 
conductance (very similar to the Zeeman effect in $S=1/2$ quantum dots). This
effect is quite clearly observed in the data of Parks {\it et al.}~\cite{ParksUS}.

We finally note that some of the data with semiconducting quantum dots with 
two electrons \cite{Kogan,Granger} are likely to display some degree of spin-orbit 
effects, although this is not yet fully clarified (see Ref.~\cite{Tagliacozzo}
for a recent discussion).

\section{Conclusion}

We have shown in this review that several striking phenomena in two-electron 
quantum dots (gate-dependent Hund's rule, singlet-triplet unscreening quantum phase 
transition and underscreened Kondo effect) can be explained in a coherent picture at 
the light of the experimental data obtained from a molecular transistor. 
Extensive Numerical Renormalization Group simulations as well as out-of-equilibrium 
diagrammatic calculations were used to confirm quasi-quantitatively these observations.
We argued that many of the observed features can be considered as generic for
molecular devices, in the absence of strong spin-orbit effects.
One crucial and common ingredient is the presence of a single active screening channel 
in the accessible temperature range, which is granted by the asymmetric hybridization 
between orbital and leads. This is expected to be satisfied in most devices,
and indeed gate-dependence of the magnetic excitations has been reported in {\it all
kinds} of quantum dot systems: semi-conducting devices \cite{Kogan}, carbon nanotubes
\cite{Quay,Holm,Delattre}, several kinds of molecules \cite{Roch,Osorio}.
The singlet-triplet quantum phase transition \cite{Roch} imposes however a further
experimental requirement, namely a small bare singlet-triplet splitting, in order to allow the 
gating effect to overcome it.
Recent theoretical work~\cite{Logan} has also confirmed a greater generality of
this physics than discussed here.
We thus conclude that the unscreening quantum phase transition and the underscreened 
Kondo effect constitute quite general paradigms to be taken for two-electron quantum 
dots.

\ack

We thank E. Eyraud, D. Lepoittevin, R. Haettel, C. Hoarau and V. Reita for their
technical contributions, E. Bonet,
T. Crozes and T. Fournier for lithography development, J. Paaske, C. Winkelmann
and C. Thirion for useful discussions, and especially T. Costi and V. Bouchiat
for
their collaboration on this topic.
Samples were fabricated in the NANOFAB facility of the N\'eel Institute.
This work is partially financed by ANR-PNANO projects MolSpintronics No.
ANR-06-NANO-27 and MolNanoSpin No. ANR-08-NANO-002, by the ERC Advanced Grant
MolNanoSpin n°226558, by the STEP MolSpinQIP, and by the PICT R1776 of
ANPCyT (Argentina).


\begin{thebibliography}{100}

\bibitem{Datta} S. Datta, ``Electronic Transport in Mesoscopic Systems'' 
(Cambridge University Press, Cambridge, England, 1997).
\bibitem{DiVentra} M. Di Ventra, ``Electrical Transport in Nanoscale Systems''
(Cambridge University Press, Cambridge, England, 2008).

\bibitem{Hanson} R. Hanson, L. P. Kouwenhoven, J. R. Petta, S. Tarucha 
and L. M.  K. Vandersypen, Rev. Mod. Phys. {\bf 79}, 1217 (2007).

\bibitem{Hewson} A. C. Hewson, ``The Kondo Problem to Heavy Fermions'', 
(Cambridge University Press, Cambridge, 1993).
\bibitem{PustilnikGlazman} M. Pustilnik and L. Glazman, J. Phys. Cond. Mat., {\bf 16},
R513 (2004).
\bibitem{GrobisReview}  M. Grobis, I. G. Rau, R. M. Potok and D. Goldhaber-Gordon, 
``Kondo Effect in Mesoscopic Quantum Dots'', Handbook of Magnetism and Magnetic 
Materials (Wiley, 2007).

\bibitem{GlazmanRaikh} L. I. Glazman and M. E. Raikh, JETP Lett. {\bf 47}, 452
(1988).
\bibitem{NgLee} T. K. Ng and P. A. Lee, Phys. Rev. Lett. {\bf 61}, 1768 (1988).

\bibitem{GoldhaberGordon} D. Goldhaber-Gordon, H. Shtrikman, D. Mahalu, 
D. Abusch-Magder, U. Meirav, and M. A. Kastner, Nature {\bf 391}, 156 (1998).
\bibitem{Cronenwett} S. M. Cronenwett, T. H. Oosterkamp 
and L. P. Kouwenhoven, Science {\bf 281}, 540 (1998).

\bibitem{Jespersen2006} T. S. Jespersen, M. Aagesen, C. Sorensen,
P. E. Lindelof, and J. Nygard,  Phys. Rev. B {\bf 74}, 233304 (2006).

\bibitem{Liang} W. Liang, M. P. Shores, M. Bockrath, J. R.  Long, and
H. Park, Nature {\bf 417}, 725 (2002).
\bibitem{ParksKondo} J. J. Parks, A. R. Champagne, G. R. Hutchison, S.
Flores-Torres, H. D. Abruna, and D. C. Ralph, Phys. Rev. Lett. {\bf 99}, 
026601 (2007).

\bibitem{NatelsonReview} G. D. Scott and D. Natelson, ACS Nano {\bf 4}, 3560
(2010).

\bibitem{Wilson} H.R. Krishna-murthy, J.W. Wilkins, and K.G. Wilson Phys. Rev.
B {\bf 21}, 1003 (1980).
\bibitem{BullaRMP} R. Bulla, T. A. Costi and T. Pruschke,
Rev. Mod. Phys. {\bf 80}, 395 (2008).

\bibitem{NozieresBlandin} P. Nozi\`eres and A. Blandin, J. Phys. (Paris) {\bf 41}, 193 (1980).
\bibitem{RochUS} N. Roch, S. Florens, T. A. Costi, W. Wernsdorfer and F.
Balestro, Phys. Rev. Lett. \textbf{103}, 197202 (2009).
\bibitem{ParksUS} J. J. Parks, A. R. Champagne, T. A. Costi, W. W. Shum, A. N.
Pasupathy, E. Neuscamman, S. Flores-Torres, P. S. Cornaglia, A. A. Aligia and C. A.
Balseiro, G. K.-L. Chan, H. D. Abru\~na and D. C. Ralph, Science {\bf 328}, 1370 (2010).



\bibitem{Nygard} J. Nygard, D. H. Cobden, P. E. Lindelof, Nature {\bf 408}, 342 (2000).
\bibitem{PustilnikAvishai} M. Pustilnik, Y. Avishai and K. Kikoin, 
Phys. Rev. Lett {\bf 84}, 1756 (2000).

\bibitem{Sasaki} S. Sasaki, S. De Franceschi, J. M. Elzerman, W. G. Van Der
Wiel, M. Eto, S. Tarucha and L. P. Kouwenhoven, Nature {\bf 405}, 764 (2000).
\bibitem{STKondo_Eto} M. Eto and Y. V. Nazarov,  Phys. Rev. Lett. {\bf 85}, 1306
(2000).
\bibitem{STKondo_Pustilnik1} M. Pustilnik and L. I. Glazman, Phys. Rev. Lett. 
{\bf 85}, 2993 (2000).
\bibitem{STKondo_Pustilnik2} M. Pustilnik and L. I. Glazman, Phys. Rev. B
{\bf 64}, 045328 (2001).


\bibitem{Allub} R. Allub and A. A. Aligia, Phys. Rev. B \textbf{52}, 7987 (1995).
\bibitem{VojtaBulla} M. Vojta, R. Bulla and W. Hofstetter, Phys. Rev. B \textbf{65}, 140405 (2002).
\bibitem{HofstetterSchoeller} W. Hofstetter and H. Schoeller, Phys. Rev. Lett.  \textbf{88}, 016803 (2002).
\bibitem{PustilnikSingletTriplet} M. Pustilnik, L. I. Glazman and W. Hofstetter, Phys.  Rev. B \textbf{68}, 161303 (2003).
\bibitem{HofstetterZarand} W. Hofstetter and G. Zarand, Phys. Rev. B {\bf 69}, 235301 (2004).
\bibitem{ZitkoSTMultidot} R. Zitko and J. Bonca, Phys. Rev. B {\bf 74}, 045312 (2006). 
\bibitem{ZitkoSTTransport} R. Zitko and J. Bonca, Phys. Rev. B {\bf 76}, 241305(R) (2007).
\bibitem{ZarandST} B. Horvath, B. Lazarovits and G. Zarand, Phys. Rev. B {\bf 82}, 165129 (2010).
\bibitem{Roch} N. Roch, S. Florens, V. Bouchiat, W. Wernsdorfer and F. Balestro, Nature \textbf{453}, 633 (2008).
\bibitem{VanDerWiel} W. G. Van der Wiel, S. De Franceschi, J. M. Elzerman, 
S. Tarucha, L. P. Kouwenhoven, J. Motohisa, F. Nakajima and T. Fukui, Phys. Rev. Lett. 
{\bf 88}, 126803 (2002).
\bibitem{Kogan} A. Kogan, G. Granger, M. A. Kastner, D. Goldhaber-Gordon, and H.
Shtrikman , Phys. Rev. B \textbf{67}, 113309 (2003).

\bibitem{PustilnikBorda} M. Pustilnik and L. Borda, Phys. Rev. B {\bf 73}, 201301 (2006).
\bibitem{Logan} D. E. Logan, C. J. Wright and M. R. Galpin, Phys. Rev. B \textbf{80}, 125117 (2009).

\bibitem{Paaske} J. Paaske, A. Rosch, P. Woelfle, N. Mason, C. M. Marcus, and J.
Nygard, Nature Physics \textbf{2}, 460 (2006).
\bibitem{RochJLTP} N. Roch, S. Florens, V. Bouchiat, W.  Wernsdorfer and F. Balestro,
J. Low Temp. Phys. {\bf 153} 350 (2008).

\bibitem{DeFranceschi} G. Katsaros, P. Spathis, M. Stoffel, F. Fournel, M. Mongillo, 
V. Bouchiat, F. Lefloch, A. Rastelli, O. G. Schmidt and S. De Franceschi, 
Nature Nanotechnology {\bf 5}, 458 (2010).
\bibitem{Jespersen} T. S. Jespersen, K. Grove-Rasmussen, J. Paaske, K.
Muraki, T. Fujisawa, J. Nyg\o{a}rd and K. Flensberg, Nature Phys. {\bf 7}, 348 (2011).
\bibitem{ZitkoSO} R. Zitko, R. Peters and Th. Pruschke, Phys. Rev. B {\bf 78}, 224404 (2008). 
\bibitem{Galpin} M. R. Galpin, F. W. Jayatilaka, D. E. Logan and
F. B. Anders, Phys. Rev. B {\bf 81}, 075437 (2010).
\bibitem{Tagliacozzo} P. Lucignano, M. Fabrizio and A. Tagliacozzo,
 Phys. Rev. B 82, 161306(R) (2010).
\bibitem{Cornaglia} P. S. Cornaglia, P. Roura Bas, A. A. Aligia and C. A.
Balseiro, EuroPhys. Lett. {\bf 93}, 47005 (2011). 
\bibitem{Andergassen} S. Grap, S. Andergassen, J. Paaske, and V. Meden,
Phys. Rev. B {\bf 83}, 115115 (2011).
\bibitem{PaaskeSO} J. Paaske, A. Andersen and K. Flensberg, Phys. 
Rev. B {\bf 82}, 081309(R) (2010).

\bibitem{Jarillo} P. Jarillo-Herrero, J. Kong, H. S. J. Van Der Zant, C. Dekker,
L. P. Kouwenhoven and S. De Franceschi, Nature {\bf 434}, 484 (2005).
\bibitem{SU4Choi} M. S. Choi, R. Lopez, and R. Aguado, Phys. Rev. Lett. {\bf 95}
067204 (2005).

\bibitem{Chung2} C.-H. Chung, G. Zarand, and P. W\"olfle, Phys. Rev. B {\bf 77}, 
035120 (2008).

\bibitem{VarmaJones} B. A. Jones, C. M. Varma, and J. W. Wilkins, 
Phys. Rev. Lett. {\bf 61}, 125 (1988).
\bibitem{Chung} G. Zarand, C.-H. Chung, P. Simon, M. Vojta
Phys. Rev. Lett. {\bf 97}, 166802 (2006).
\bibitem{Vavilov} M. G. Vavilov and L. I. Glazman Phys. Rev. Lett. {\bf 94}, 086805 (2005).
\bibitem{Craig} N. J. Craig, J. M. Taylor, E. A. Lester, C. M. Marcus, M. P. Hanson
and A. C. Gossard, Science {\bf 304}, 565 (2004).

\bibitem{KoenigGefen} J. K\"onig and Y. Gefen, Phys. Rev. B {\bf 71}, 201308 (2005).

\bibitem{LeoFabrizio} L. De Leo and M. Fabrizio, Phys. Rev. Lett. {\bf 94}, 236401 (2005).
\bibitem{Lazarovits} B. Lazarovits, P. Simon, G. Zarand and L. Szunyogh,
Phys. Rev. Lett. {\bf 95}, 077202 (2005).
\bibitem{Ingersent} K. Ingersent, A. W. Ludwig and I. Affleck
Phys. Rev. Lett. {\bf 95}, 257204 (2005).
\bibitem{AligiaTrimer} A. A. Aligia, Phys. Rev. Lett. {\bf 96}, 096804 (2006).
\bibitem{MitchellLogan} A. K. Mitchell and D. E. Logan, Phys. Rev. B {\bf 81}, 075126 (2010).
\bibitem{Ferrero} M. Ferrero, L. De Leo, P. Lecheminant and M. Fabrizio,
J. Phys.: Condens. Matter {\bf 19} 433201 (2007).


\bibitem{Quay} C. H. L. Quay, J. Cumings, S. Gamble, R. de Picciotto, H.
Kataura, and D. Goldhaber-Gordon, Phys. Rev. B {\bf 76}, 245311 (2007).
\bibitem{Delattre} T. Delattre, C. Feuillet-Palma, L.G. Herrmann, P. Morfin,
J.-M. Berroir, G. Fève, B. Plaçais, D.C. Glattli, M.-S. Choi, C. Mora, and T.
Kontos, Nature Physics \textbf{5}, 208 (2009).

\bibitem{Holm} J. V. Holm, H. I. Jorgensen, K. Grove-Rasmussen, J.  Paaske, 
K. Flensberg and P. E. Lindelof Phys. Rev. B \textbf{77}, 161406(R) (2008).
\bibitem{Hauptmann} J. R. Hauptmann, J. Paaske and P. E. Lindelhof, Nature 
Physics \textbf{4} 373 (2008).
\bibitem{Osorio} E. A. Osorio, K. Moth-Poulsen, H. S. J. van der Zant,
J. Paaske, P. Hedegard, K. Flensberg, J. Bendix, T. Bjornholm, Nano Lett.
{\bf 10}, 105 (2010).

\bibitem{Aligia1} P. Roura-Bas and A. A. Aligia, Phys. Rev. B \textbf{80}, 035308 (2009).
\bibitem{Aligia2} P. Roura-Bas and A. A. Aligia, J. Phys. Cond. Mat. 
\textbf{22}, 025602 (2010).

\bibitem{Freyn2} A. Freyn and S. Florens, in preparation.

\bibitem{RochPhys} N. Roch, C.B. Winkelmann, S. Florens, V. Bouchiat, W.
Wernsdorfer and F. Balestro, Phys. Stat. Sol. B {\bf 245}, 1994 (2008).

\bibitem{Schmid} J. Schmid, J. Weis, K. Eberl and K. v. Klitzing, Phys. Rev.
Lett. {\bf 84}, 5824 (2000). 
\bibitem{Granger} G. Granger, M. A. Kastner, Iuliana Radu, M. P. Hanson 
and A. C. Gossard, Phys. Rev. B {\bf 72}, 165309 (2005).

\bibitem{Pruschke} R. Peters and T. Pruschke, New J. Phys. \textbf{8}, 127
(2006).

\bibitem{Anders} F. B. Anders, Phys. Rev. Lett. {\bf 101}, 066804 (2008).

\bibitem{Freyn} A. Freyn and S. Florens, Phys. Rev. B {\bf 79}, 121102(R) (2009).

\bibitem{MeirWingreen} Y. Meir and N. S. Wingreen, Phys. Rev. Lett. {\bf 68}, 2512
(1992).

\end{thebibliography}
\end{document}